\documentclass[%
reprint,
 amsmath,amssymb,
 aps,
prb,
]{revtex4-2}

\usepackage{longtable}
\usepackage{pgfplots}
\usepackage{tabularx}
\usepackage[colorlinks=true,citecolor=blue, linkcolor=red, urlcolor=blue]{hyperref}
\usepackage{xcolor}
\usepackage{graphicx}
\usepackage{amssymb}
\usepackage{amsmath}
\usepackage{amsfonts}
\usepackage{mathtools}
\usepackage{comment}
\usepackage{tikz}
\usepackage{graphicx}
\usepackage{dcolumn}
\usepackage{bm}

\begin{document}


\def\titlename{Nonlinear optical responses in nodal line semimetals}
\title{\titlename}
\def\authornames{Omid Tavakol, Yong Baek Kim}
\author{Omid Tavakol}
\author{Yong Baek Kim}


\def\affiliations{Department of Physics, University of Toronto, Toronto, Ontario, M5S 1A7, Canada}
\affiliation{\affiliations}

\begin{abstract}
Shift current and second harmonic generation (SHG) are nonlinear optical responses that are often used for photovoltaic effect and the detection of subtle symmetry breaking patterns in materials. It has been recently shown that topological semimetals such as Weyl semimetals with broken inversion symmetry can generate large nonlinear optical responses. In this paper, we investigate both shift current and SHG in a class of topological nodal line semimetal (NLSM) systems with broken inversion symmetry, which may arise intrinsically or via an applied electric field. 
It is shown that certain classes of NLSM systems offer more singular shift current and SHG in comparison to the well-known Weyl semimetal systems.


\end{abstract}


\maketitle

\section{Introduction}
Nonlinear optical responses of quantum materials offer deeper understanding of many-body excitation spectra\cite{nonlinearbook}, which may help us build more useful optical devices\cite{MooreSHG}. Recent studies show that the bulk photovoltaic effect or the shift current, which is a nonlinear optical response producing a net photocurrent in non-centrosymmetric
materials\cite{defshift1,defshift2,introshift1,introshift2,introshift3}, is a promising substitute for conventional solar cells with {\it pn} junctions\cite{MooreShift}. Further, the SHG is another second order nonlinear optical response, where the material absorbs a light with frequency $\omega$ and produces a light with doubled frequency $2\omega$,\cite{sipe,introSHG,SHGapp3,PhysRevB.98.165113}. The symmetry structure of SHG tensor provides important information about the subtle broken symmetries in materials\cite{SHGapp,appSHG2}, which may not be obvious in linear response probes. Although the basic principles of these nonlinear responses are well understood\cite{sipe,mainshift,Topological_Morimoto}, it is highly valuable to investigate materials that may provide strong nonlinear optical responses.
 
Recently, it has been shown that topological semi-metals such as Weyl semi-metals\cite{RevModPhys.90.015001}, where the valence and conduction bands cross each other in momentum space at a discrete set of points, can produce strong nonlinear optical responses\cite{strongresweyl1,strongweyl2,experminet,Osterhoudt2019}. These responses are intimately related to the topology of the wavefunction near the band crossings points known as Weyl points\cite{Berry_NLR,BerryNLR2,Berry_NLR3,Berry_NLR4}. For example, a strong nonlinear optical response was observed in a Weyl semi-metal with broken inversion symmetry, TaAs\cite{experminet,shiftweyl,MooreSHG,PhysRevLett.122.197401}. 

\begin{figure}[h]
    \centering
    \includegraphics[width=0.3\textwidth]{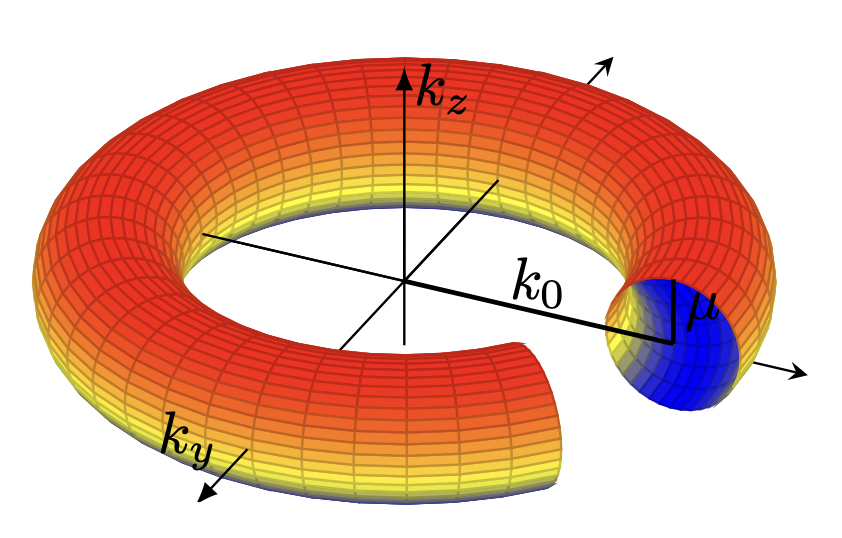}
    \caption{The Fermi surface of a NLSM. The radius of the torus is considered to be $k_0$ and $k_0 > \mu$, where $\mu$ is the chemical potential. }
    \label{fig:my_label}
\end{figure}

In this paper, we investigate nonlinear optical responses in a class of topological nodal line semi-metals (NLSM)\cite{PhysRevB.92.081201,Fang2016,PhysRevB.92.045126}. The crossing points between conduction and valence bands form a closed loop in the momentum space and, upon doping, a Fermi surface with the topology of a torus arises as shown in FIG.~\ref{fig:my_label}.   
There exist a number of materials that show NLSM behavior such as $\text{HgCr}_2\text{Se}_4$\cite{HgCr}, $\text{Cu}_3\text{PdN}$\cite{CuPdN}, $\text{Cu}_3\text{ZnN}$\cite{PhysRevLett.115.036806}, $\text{Ca}_3\text{P}_2$\cite{PhysRevB.93.205132},  ZrSiS\cite{ZrSiS} and SrI$\text{O}_3$\cite{SrIo}. The NLSM exhibits various exotic properties due to the nontrivial structure of the Fermi surface, which includes parity anomaly\cite{ShiftNLSM} and giant nonlinear response in the presence of magnetic fields\cite{sinha}. 

The most important quantities that can affect nonlinear optical conductivity are energy dispersion and topology of the Bloch wavefunction.
In this work, we consider a class of NLSM models with different dispersion and study the behaviors of shift current and SHG. Since these nonlinear responses arise when the inversion symmetry is broken, we consider the systems where the inversion symmetry is broken either intrinsically or via an external electric field. 
We first consider the NLSM where the inversion symmetry is broken by an external electric field. It is found that SHG in the NLSM can have a more singular response in comparison to the Weyl semimetal, {\it i.e.} 
$\sigma^{\text{NLSM}}_{\text{SHG}} \sim \sigma^{\text{weyl}}_{\text{SHG}}(\frac{vk_0}{\mu}) $ in low doping limit $\mu\rightarrow 0$ ($vk_0$ is an energy scale associated with the NLSM). Furthermore, considering the NLSM systems with intrinsically broken inversion symmetry, we show that certain class of NLSM can have stronger shift current (and similarly SHG) responses,  $\sigma^{zzz}_{\text{shift,NLSM}}(0;\omega,-\omega) \propto (e^3 / \hbar^2) \omega^{-3/2}$, in comparison to the case of Weyl semimetals $\sigma^{zzz}_{\text{shift,Weyl}}(0;\omega,-\omega) \propto (e^3 / \hbar^2)\omega^{-1}$ in the low frequency limit. We explain below how these singular behaviors arise from the peculiar structure of the Fermi surface and the Berry curvature of the wavefunction in NLSM.

The rest of the paper is organized as follows. In section II, we discuss the nonlinear optical response functions in different gauge choices. In particular, the relations between the expressions of shift current and SHG in different gauge choices are discussed. In section III, we consider a NLSM, where the inversion is broken by an external dc electric field. We show the behavior of SHG in this system. 
In section IV, we introduce two classes of NLSM systems with intrinsically broken inversion symmetry. We study both shift current and SHG and present their behaviors. In section V, we discuss the implications of our results.

\section{nonlinear optical responses in length and velocity gauges}

\subsection{Length and Velocity Gauges}

We first introduce the general expressions of the shift current and SHG responses in the length and velocity gauges. The purpose is to show the direct relations between these expressions in different gauge choices. 

For simplicity, we start with a single particle Hamiltonian with a quadratic dispersion, 
$H({\bf p},{\bf r})=H_0+V_{int}({\bf r})$ where $H_0={\bf p}^2/2m$ and $V_{int}({\bf r})$ is the interaction with an electromagnetic field.
In the presence of an electromagnetic field, $V_{int}({\bf r})$ is obtained from the minimal coupling, ${\bf p}\rightarrow {\bf p}-e{\bf A}$, which leads to

\begin{equation}
    H= 
    \frac{{\bf p}^2}{2m}-\frac{e}{2m}\big({\bf A}({\bf r},t)\cdot {\bf p} + {\bf p} \cdot {\bf A}({\bf r},t)\big)+\frac{e^2}{2m}{\bf A}^2({\bf r},t)-e\phi({\bf r},t)\,,
    \label{Hgauge}
\end{equation}
where ${\bf A}$ and $\phi$ are vector and scalar potentials of the electromagnetic field.
In the absence of any static source, we can assume that $\phi(r,t)=0$ and also in the dipole approximation, we can take the vector potential to be independent of position ${\bf A}(r,t)={\bf A}(t)$.
In finding the second order response, the quadratic term in the gauge field can make the calculation much more difficult. As a result, it would be more efficient to cancel this term by fixing a gauge. The first gauge that can be used is the length gauge. The freedom of gauge transformation can be written as follows 
\begin{align}
   & {\bf A}\rightarrow {\bf A'}={\bf A}+{\bf \nabla} \Lambda \ , \\\nonumber&
   \phi\rightarrow\phi'=\phi-\partial_t\Lambda \ , \\\nonumber&
   \psi\rightarrow\psi'=e^{-ie\Lambda/\hbar}\psi\,,
   \label{gaugetransformation}
\end{align}
where 
$\psi$ is the matter field or wavefunction. $A'$, $\phi'$ and  $\psi'$ are transformed vector potential, scalar potential, and matter field respectively and $\Lambda$ is an arbitrary function. 
By choosing $\Lambda=-{\bf A}(t)\cdot {\bf r}$, we get ${\bf A'}=0$ and $\phi'=\partial_t {\bf A}(t)\cdot{\bf r}=-{\bf E}(t) \cdot {\bf r}$.
Then, the Hamiltonian would be 
\begin{equation}
    H_L=\frac{\bf p^2}{2m}+e{\bf E}(t)\cdot {\bf r}\,,
    \label{HLength}
\end{equation}
where the subscript $L$ denotes that the Hamiltonian is written in the length gauge. ${\bf E}(t)$ is the electric field and $e {\bf E}(t)\cdot {\bf r}$ term can be interpreted as an interaction between the electric field and the polarization $\hat{\bf P}_{tot}=-e\hat{\bf r}$.

In the systems with multi-band structure, there exist two kinds of polarizations due to the interband   and intraband  interactions, which we denote them as ${\bf P}_{e}$ and ${\bf P}_{i}$, respectively. 
The second quantized forms of the polarizations can be written as follows 
\begin{equation}
    \hat{\bf P}_{e}=e \int [dk] \sum_{nm}{\bf r}_{nm}a^{\dagger}_{n}(k)a_{m}(k)\,,
    \label{Pe}
\end{equation}
\begin{equation}
    \hat{\bf P}_{i}=e \int [dk] \sum_{n}(a^{\dagger}_{n}(k)\partial_{\bf k} a_{n}(k)-i\bm{ \xi}_{nn}a^{\dagger}_{n}(k)a_{n}(k))\,,
    \label{Pi}
\end{equation}
where $[dk] = \frac{d^3 k}{(2\pi)^3}$. Here, $a^{\dagger}$ and $a$ are fermionic creation and annihilation operators. $\bm{\xi}_{nn}$ is the Berry connection of the $n$th band and ${\bf r}_{nm}$ is the expectation value of the position operator between $n$ and $m$ states, $\langle n | {\bf r} | m \rangle$, which can be written in the following form
\begin{align}
    \bm{\xi}_{nm}=i u_n(k)^{*}\partial_{\bf k}u_m(k) ,\\\nonumber 
    {\bf r}_{nm}=\bm{\xi}_{nm}\,\,\,\,\,{\rm when}\,\,\,\,n\neq m\,,
    \label{rnm}
\end{align}
where $u_n(k)$ is the periodic part of the Bloch wavefunction.

On the other hand, for the velocity gauge, we can choose $\Lambda=-\frac{e}{2m}\int_{-\infty}^{t}{\bf A}(t')\cdot {\bf A}(t')dt'$. Thus, we find ${\bf A'}={\bf A}(t)$ and $\phi'=\frac{e}{2m}{\bf A}(t)\cdot {\bf A}(t)$. Then, the Hamiltonian would be 
\begin{equation}
    H_V=\frac{{\bf p}^2}{2m}+\frac{e}{m}{\bf A}(t)\cdot{\bf p}\,,
    \label{Hvelocity}
\end{equation}
where the subscript $V$ denotes that the Hamiltonian is written in the velocity gauge.
Note that the Hamiltonians in both gauge choices are linear in the external gauge field, making the calculation easier. 

To complete the discussion, the description above can be generalized to the systems with a general dispersion $H_0 = \varepsilon ({\bf p})$ and $H ({\bf p},{\bf r}) = H_0 + V_{int}({\bf r})$. In the presence of an external gauge field, one can use the minimal substitution ${\bf p}\rightarrow{\bf p}-e{\bf A}$. Then the Hamiltonian up to the second order in gauge field can be written in the following form
\begin{equation}
H = \varepsilon({\bf p})-e \nabla_{p} \varepsilon({\bf p})\cdot{\bf A}+\frac{e^2}{2} \nabla_{p}^2 \varepsilon({\bf p}) ({\bf A}\cdot {\bf A})+O({\bf A}^3)-e\phi \ .
\label{Hgen}
\end{equation}
Again, by choosing $\Lambda=-{\bf A}(t)\cdot {\bf r}$, we get ${\bf A'}=0$ and $\phi'=\partial_t {\bf A}(t) \cdot {\bf r}=-{\bf E}(t) \cdot {\bf r}$ in the length gauge. By making the vector potential vanish, the only term contributing to the Hamiltonian is the scalar potential. Thus, the Hamiltonian in the presence of an external electric field in the length gauge can be written in the following form
\begin{equation}
H_L({\bf p},r)=\varepsilon ({\bf p})+e {\bf E}(t) \cdot {\bf r} \ .
\end{equation}
Similar consideration applies to the Hamiltonian with a general dispersion in the velocity gauge.

\subsection{Shift Current and Second Harmonic Generation}
In the presence of an electric field $\bf{E}$, the response of the system can be written in term of an expansion of the current, ${\bf J}$, as a function of the electric field,
\begin{equation}
J^a(\omega)=\sigma^{ab}(\omega)E^b(\omega)+\sigma^{abc}(\omega;\omega_1,\omega_2)E^{b}(\omega_1)E^{c}(\omega_2)+..., 
\label{gencurrent}
\end{equation}
where $a,b,c \in \{x,y,z\}$, $\omega=\omega_1+\omega_2$, and $E(\omega_1)$ is the Fourier component of the time-dependent electric field. In this manner, $\sigma^{ab}(\omega)$ is the linear response and $\sigma^{abc}(\omega;\omega_1,\omega_2)$ is the second order nonlinear response to the incident light. There are a number of nonlinear optical processes that can occur in the presence of electric field, which include SHG, sum and difference frequency generation, and shift current that is also known as bulk photovoltaic effects (BPVE). These second order responses can be observed in systems with broken inversion symmetry. In this work, we only focus on the SHG and shift current.

\subsubsection{Shift Current}
In systems with broken inversion symmetry, the separation of centers of charge in valence and conduction bands can produce the net {\it dc} current, known as the shift current\cite{mainshift}. In the length gauge formulation, if the light is polarized in $b$ direction, the shift current can be written in the following form 
\begin{align}
   \label{shiftlength}
    \sigma^{abb}_{\rm shift}&(0;\omega,-\omega)=\\\nonumber&
    \frac{\pi e^3}{\hbar^2}\sum_{nm}\int_{BZ}[dk]|r^{b}_{nm}(k)|^2R^{a}_{nmb}(k)f_{nm}\delta(\omega_{nm}-\omega) \,,
\end{align}
where the sum is over the band index.
Here, $a,b\in(x,y,z)$ are spatial coordinates and $f_{nm}=f(\epsilon_n)-f(\epsilon_m)$, where $f(\epsilon)$ is the Fermi distribution function. Note that $\epsilon_{n}=\hbar \omega_{n}$ is the eigenvalue of the Bloch Hamiltonian $H(k)$ and $\omega_{nm}=\omega_n-\omega_m$.  Moreover, $f_{nm}|r^{b}_{nm}(k)|^2$ is related to the transition intensity and $R^{a}_{nmb}(k)$ denotes the separation of centers of charge (in the valence and conduction bands) in the real space, which is known as the shift vector
\begin{equation}
   R^{a}_{nmb}(k) =-\frac{\partial \phi^b_{nm}}{\partial k_{a}}+\xi_{nn}^{a}-\xi_{mm}^{a}\,.
   \label{shiftvec}
\end{equation}
Here, $\phi^b_{nm}={\rm Im}[\ln(r^b_{nm})]$ is the phase of the matrix element of the position operator.

On the other hand, in the velocity gauge, the shift conductivity can be written as follows
\begin{align}
    \sigma_{{\rm shift},V}^{abb}&(0;\omega,-\omega) =\\\nonumber&
    \sum_{nm}\frac{\pi e^3}{\hbar^2\omega^2}\int [dk]\delta(\epsilon_n-\epsilon_m+\hbar \omega)|v^b_{nm}|^2 f_{nm} R^{a}_{nmb}\,,
    \label{shiftvelocity}
\end{align}
where $v^b_{nm}=\big<n|\frac{\partial H}{\partial k_b}| m\big>$ is the matrix element of the velocity operator.
The shift conductivity in velocity gauge can be related to that of the length gauge by using the identity $r^b_{nm}=-\frac{v^b_{nm}}{\epsilon_n-\epsilon_m}$, which leads to the same result as in the length gauge, 
\begin{align}
    \sigma_{{\rm shift},V}^{abb}&(0;\omega,-\omega)
    =\\\nonumber&
    \sum_{nm}\frac{\pi e^3}{\hbar^2}\int [dk]\delta(\epsilon_n-\epsilon_m+\hbar \omega)|r^b_{nm}|^2f_{nm}R^{a}_{nmb}\,.
\end{align}

\subsubsection{Second Harmonic Generation}

In the SHG process, systems with broken inversion symmetry can absorb a light with frequency $\omega$ and produce a light with doubled frequency $2\omega$. The corresponding nonlinear current can be written as
$J^{a}(2\omega)=\sigma^{abc}(2\omega;\omega,\omega)E^{b}(\omega)E^{c}(\omega)$.
Here, the SHG response tensor in the velocity gauge
can be defined as follows
\begin{equation}
    \sigma^{abc}_{SHG}(\omega)=\sigma^{abc}_{2p,I}(\omega)+\sigma^{abc}_{2p,II}(\omega)+\sigma^{abc}_{1p,I}(\omega)+\sigma^{abc}_{1p,II}(\omega)\,,
    \label{SHGgen}
\end{equation}
where 
\widetext
\begin{align}
\label{SHG1}
    &\sigma^{abc}_{2p,I}(\omega)=\sum_{nm}\frac{e^3}{2\hbar^2\omega^2}\int[dk]v^{a}_{mn}w^{bc}_{nm}f_{mn}R_{\gamma}(2\omega-\omega_{nm})\,,\\&
    \label{SHG2}
    \sigma^{abc}_{2p,II}(\omega)=\sum_{nmp}\frac{e^3}{2\hbar^2\omega^2}\int[dk]\frac{2v^{a}_{mn}[v^{b}_{np}v^{c}_{pm}]_{+}}{\omega_{mp}+\omega_{np}}f_{mn}R_{\gamma}(2\omega-\omega_{nm})\,,\\&
    \label{SHG3}
    \sigma^{abc}_{1p,I}(\omega)=\sum_{nm}\frac{e^3}{2\hbar^2\omega^2}\int[dk](w^{ab}_{mn}v^{c}_{nm}+w^{ac}_{mn}v^{b}_{nm})f_{mn}R_{\gamma}(\omega-\omega_{nm})\,,\\&
     \sigma^{abc}_{1p,II}(\omega)=\sum_{nmp}\frac{e^3}{2\hbar^2\omega^2}\int[dk]\frac{v^{a}_{mn}[v^{b}_{np}v^{c}_{pm}]_{+}}{\omega_{pm}+\omega_{pn}}(f_{mp}R_{\gamma}(\omega-\omega_{pm})-f_{np}R_{\gamma}(\omega-\omega_{np}))\,.
     \label{SHG4}
\end{align}
\endwidetext
Here, $w^{ab}=(1/\hbar)\partial_{k_a}\partial_{k_b}H$, $[v^{b}_{np}v^{c}_{pm}]_{+}=v^b_{np}v^c_{pm}+v^c_{np}v^b_{pm}$ and $R_{\gamma}(x)=1/(x-i\gamma)$, where $\gamma$ is related to the decay rate. SHG consists of four different processes Eq.~(\ref{SHG1}-\ref{SHG4}). The first two equations Eq.~(\ref{SHG1}) and Eq.~(\ref{SHG2}) which are denoted by index 2p show the contributions of two photon resonance. On the other hand the Eq.~(\ref{SHG3}) and Eq.~(\ref{SHG4}) denoted by 1p show the contributions of one photon resonance.

\section{Nodal line semimetal with inversion-breaking external electric field}

In this section, we start with an inversion symmetric model for nodal line semimetals. Then, in order to find a finite SHG response, we break the inversion symmetry by applying an external electric field. 
We find the analytic expression for SHG in this case.

Let us consider the following Hamiltonian \cite{Rhim2016} 
\begin{equation}
    H(k)=v(\sqrt{k_x^2+k_y^2}-k_0)\tau_x+v k_z \tau_z\,,
    \label{NLSMclass}
\end{equation}
where $\tau_{A}$ are the Pauli matrices acting on the orbital space, where $A \in \{x,y,z\}$. Here $k_0$ is the radius of the nodal line in the momentum space.
The location of the zero energy nodal line spectrum is given by $\sqrt{k_x^2+k_y^2}=k_0$ and $k_z=0$. This Hamiltonian preserves both time reversal and inversion symmetry as it commutes with the parity operator $\mathcal{P}=\tau_x \otimes (k\rightarrow -k)$ and the time reversal operator $T=K\otimes (k \rightarrow -k)$, where $K$ is complex conjugate operator. 

Now we introduce an external electric field to break the inversion symmetry. 
In the systems with inversion symmetry, the second order nonlinear response would vanish and only odd responses would contribute to the current density. However, in the presence of an external {\it dc} electric field, the distribution function of electrons breaks the inversion symmetry and this gives rise to the second order nonlinear response such as SHG.

In the presence of an external {\it dc} electric field, we can find the distribution function using the Boltzmann equation by introducing a relaxation time $\tau$,
\begin{equation}
-\frac{e}{\hbar} {\bf E} \cdot \partial_{\bf k}f
=-\frac{\delta f}{\tau} \ .
\end{equation}
By solving the Boltzmann equation, one can obtain the distribution function for the $n$th band
\begin{equation}
f_n= f_n^{(0)}+\frac{e\tau }{\hbar}{\bf E}\cdot\partial_{\bf k} f^{(0)}_n+ \frac{e\tau }{\hbar} E^a E^b\partial_{k_{a}}f^{(0)}_n\partial_{k_{b}}f^{(0)}_n \ ,
\label{Boltzmann}
\end{equation}
where $f^0_n$ is the equilibrium distribution function.
The important contribution is the linear term in the electric field, which breaks the inversion and gives rise to the SHG. 

Let us consider a {\it dc} electric field in $z$ direction. In our model, in the low frequency limit, $\sigma_{2p,I}^{abc}$ and $\sigma_{1p,I}^{abc}$ vanish. Note that these two contributions also vanish in semimetals with a linear dispersion, such as Weyl semimetals \cite{mainshift}. The reason why these contributions vanish in Weyl seminmetals is that
the second derivative of the Hamiltonian, which enters in $\sigma_{2p,I}^{abc}$ and $\sigma_{1p,I}^{abc}$, is zero. However, in our model, these terms do not vanish explicitly, but the integral over the Fermi surface vanishes because they are odd functions of momentum.
The only important remaining terms are $\sigma_{2p,II}^{abc}$ and $\sigma_{1p,II}^{abc}$ in Eq.~(\ref{SHG2}) and Eq.~(\ref{SHG4}), respectively. Both of these terms scale similarly with respect to frequency, which is explained in the analytic expression of SHG in Appendix C.

In the case of $\mu < v k_0$, where $\mu$ is the chemical potential, we have the following form for SHG
\begin{equation}
    \sigma^{abc}_{SHG}(2\omega;\omega,\omega)=\sigma_{2p,II}^{abc}(2\omega;\omega,\omega)+\sigma_{1p,II}^{abc}(2\omega;\omega,\omega)\,,
\end{equation}
where
\begin{align}
    &\sigma_{2p,II}^{abc}(2\omega;\omega,\omega)=\frac{e^4v^2E_zk_0\tau}{h^3\mu^2}R_\gamma(2\omega-2\mu)C_{2pII}^{abc}\,,
    \label{Eq_results_SHG}\\&
    \sigma_{1p,II}^{abc}(2\omega;\omega,\omega)=\frac{e^4v^2E_zk_0\tau}{h^3\mu^2}R_\gamma(\omega-2\mu)C_{1pII}^{abc}\,.
\end{align}
Here $C_{1pII}^{abc}$ and $C_{2pII}^{abc}$ are numerical coefficients shown in Table~\ref{coeftab} in the Appendix C, and $R_{\gamma}(\omega-2\mu)=\frac{1}{\hbar \omega-2\mu-i\gamma}$. 
We assume $\gamma$ is sufficiently small so that the resonant factor $R_{\gamma}(x)$ behaves like $\delta(x)$. Thus we used an approximation  $\omega \sim 2\mu$ and $\omega \sim \mu$ for the one-photon and two-photon resonances, respectively.

In the limit $\mu \ll vk_0$, the SHG is singular as
\begin{equation}
\sigma^{nod}_{SHG}\sim k_0/\mu^2 \ ,
\label{SHGapp1}
\end{equation}
which is more divergent in comparison to the case of Weyl semimetal, $\sigma^{weyl}_{SHG}\sim1/\mu$. 
On the other hand, in the opposite limit, $vk_0\ll \mu$, we obtain 
\begin{equation}
\sigma^{nod}_{SHG}\sim 1/\mu+k_0/\mu^2+O((k_0/\mu)^2) \ .
\label{SHGapp2}
\end{equation}
Note that, in the limit $k_0\rightarrow 0$, this result reduces to the same behavior as in Weyl semimetal.

\section{nodal line semimetal with 
intrinsically broken Inversion symmetry}

In this section, we introduce two sets of Hamiltonian (Model 1 and Model 2) for NLSM with intrinsically broken inversion symmetry. We also consider tunable dispersions in these models, which lead to different behaviors in the density of states.  
For each model, we compute the shift current and SHG for different dispersion relations and we show that some choice of dispersion can enhance the shift current response.  

\subsection{Two models}


Let us first consider the following $2\times2$ Hamiltonian, which we call Model 1.
\begin{equation}
    H(k)=t_x (\sqrt{k_x^2+k_y^2}-k_0)^a\tau_x+v_y k_z \tau_y+t_z k_z^b \tau_z\,,
    \label{NLSM1}
\end{equation}
where $\tau_{A}$ are Pauli matrices acting on the orbital space, where $A \in \{x,y,z\}$. $k_0$ is the radius of the nodal line in momentum space. 
The location of the zero energy nodal line is given by
$\sqrt{k_x^2+k_y^2}=k_0$ and $k_z=0$. 

There is another class of Hamiltonian that can capture the feature of nodal line semimetals but with different scaling with respect to $x,y$ and $z$ directions\cite{PhysRevB.97.165118,PhysRevB.93.035138,PhysRevB.97.161113,ShiftNLSM,sinha}. This effective Hamiltonian, which we call Model 2, can be written in the following form
\begin{equation}
    \tilde{H}=\tilde{t}_x(k_x^2+k_y^2-k_0^2)^a\tau_x+v_y k_z\tau_y+t_z k_z^b\tau_z\,.
    \label{NLSM2}
\end{equation}

In both models, the Hamiltonian can break inversion symmetry or time-reversal symmetry depending on the exponent $b$. This can be determined by considering the commutation relation between the Hamiltonian and the parity operator $\mathcal{P}=\tau_x \otimes (k\rightarrow -k)$ and the time reversal operator $T=K\otimes (k \rightarrow -k)$, where $K$ is complex conjugate operator. In this model, the $\tau_z$ term breaks the inversion when $b$ is an even integer (but preserves time-reversal symmetry). Note that, as long as $b>1$, $\tau_z$ term can be regarded as a small perturbation in the low frequency limit. We investigate the behavior of the shift current depending on the choice of $a$ and $b$.  
\subsection{Shift conductivity}

One of the important factors determining the scaling behavior of the shift current is the joint density of states, $\rho(\epsilon)$. This, as well as the shift-vector and transition intensity, determine the scaling behavior of the shift conductivity in the low-frequency limit. In the following, we examine the contribution of each of these three quantities to the shift current in Model 1 and Model 2, respectively. The joint density of state for these two models can be written as follows. The quantity represented by the variables with the tilde sign is for the Model 2.
\begin{align}
    \label{density1}
    \rho(\omega)&=\frac{1}{V}\int[dk] \delta(\omega-\omega_{cv}(k))\\\nonumber
    &=2\pi\frac{4k_0\sqrt{\pi}}{v_y}\left (\frac{\omega}{t_x}\right )^{\frac{1}{a}}\frac{\Gamma(1+\frac{1}{2a})}{\Gamma(\frac{1+a}{2a})}\,,
\end{align}
\begin{align}
    \label{density2}
    \tilde{\rho}(\omega)&=\frac{1}{V}\int[dk] \delta(\omega-\tilde{\omega}_{cv}(k))\\\nonumber &=2\pi\frac{\sqrt{\pi}}{av_y}\left (\frac{\omega}{\tilde{t}_x} \right )^{\frac{1}{a}}\frac{\Gamma(\frac{1}{2a})}{\Gamma(\frac{1+a}{2a})}\,.
\end{align}
Here, $\hbar \omega_{cv} = \epsilon_c - \epsilon_v$,
$\hbar {\tilde \omega}_{cv} = {\tilde \epsilon_c} - {\tilde \epsilon}_v$ and 
$c,v$ represent the conduction and valence bands.
For both models, the scaling of the density of states as a function of energy is the same. However, in Model 1, the presence of a finite radius $k_0$ is crucial to find a non-zero density of state, which is not the case for Model 2.

Let us first consider Model 1. To find the relation between the joint density of state and shift conductivity in Eq.~(\ref{shiftlength}), we change the integral variable from lattice momentum to $\epsilon$, $\Omega$ which are energy and the solid angle in momentum space, respectively. Then we can find the following form for the shift conductivity   
\begin{equation}
\label{Shiftvarchange}
    \sigma_{\rm shift}^{ijj}(\omega)\\\nonumber
    =\int d\Omega \ d\epsilon\, I^{ijj}(\theta,\epsilon) f_{cv}(\epsilon) \delta (\omega_{cv} - \omega) \ ,
\end{equation}
where
\begin{equation}
  I^{ijj}(\theta,\epsilon)= |\det J| \  |r^j_{cv}(\epsilon,\theta)|^2 R^i_{cv,j}(\epsilon,\theta) \ .
  \label{Iabbshift}
\end{equation}
Here, 
$f_{cv}(\epsilon)=f(\epsilon_c)-f(\epsilon_v)$, $|\det J|$ is the determinant of the Jacobian matrix and $i,j \in \{x,y,z\}$. The exact expression of the Jacobian determinant is discussed in Appendix B, shown in Eq.~(\ref{Jacobian}). It can be schematically written as $|\det J| (\epsilon,\theta) = k_0\,\epsilon^{1/a}u_{a}(\theta)+\epsilon^{2/a}g_a(\theta)$. Here, $u_a$ and $g_a$ are functions of $\theta$ and $a$. The contribution of the first term corresponds to the joint density of state for Model 1, shown in Eq.~(\ref{density1}). This can be seen from $\int d\Omega \ |\det J| (\epsilon,\theta) \propto \rho(\epsilon)$ due to $\int d\Omega \ g_a (\theta) = 0$.
Thus, we can write the determinant of the Jacobian matrix in terms of the joint density function, $|\det J| (\epsilon,\theta) = \rho(\epsilon)h_{a}(\theta)+\epsilon^{2/a}g_a(\theta)$. Here, $h_a(\theta)$ and $u_{a}(\theta)$ differ only by a multiplicative constant.   
For the scaling behavior of the shift conductivity, the other two important factors are the transition intensity, which is proportional to the interband   element of the position operator $|r^{j}_{nm}|^2$ and the shift-vector in Eq.~(\ref{shiftvec}).

To investigate the contribution of each term, we need to choose a direction. Because of the symmetry of the Hamiltonians,
$x$ and $y$ directions behave similarly, so let us only consider $zzz$ and $zxx$ directions. Let us first consider the $zzz$ direction. In our two-band Hamiltonian models, in the low frequency limit, we find that the interband element of the position operator scales as $|r^{z}_{cv}(\epsilon,\theta)|^2 = \epsilon^{-2} h^{r,z}(\theta)$. Here, $h^{r,z}(\theta)$ is an angular part of the $|r^{z}_{cv}(\epsilon,\theta)|^2$ term for Model 1.

The shift-vector is another factor that has the topological information about the wave function. The Berry connection explicitly contributes to the shift-vector via Eq.~(\ref{shiftvec}), which is a gauge-invariant physical quantity. 
There are intraband  and interband   contributions to the shift-vector, coming from  $\xi^{z}_{nn}$ and $ \partial_{k_z}\phi^{z}_{nm}$, respectively. In our models, in the low frequency limit, both of these quantities scale as $\epsilon^{-2+b}$. Thus we can write the shift vector as $R_{cv;z}^z=\epsilon^{-2+b}h^{R,zz}_{a,b}(\theta)$. Here, $h^{R,zz}_{a,b}(\theta)$ is an angular part of the shift vector along $z$ direction in Model 1, which depends on both $a$ and $b$ exponents.

Finally, in the low frequency  limit, the simplified form of Eq.~(\ref{Shiftvarchange}) (as discussed in Appendix B, Eq.~(\ref{IJ+zzz})) can be written as the product of Jacobian determinant, transition intensity and the shift vector. By multiplying these terms we find that the integrand of shift conductivity shown in Eq.~(\ref{Shiftvarchange}) can be written as 
\begin{equation}
 I^{zzz}(\epsilon,\theta)=\frac{k_0}{\epsilon^{4-b-\frac{1}{a}}}F^{zzz}_{a,b}(\theta)+\frac{1}{\epsilon^{4-b-\frac{
    2}{a}}}G^{zzz}_{a,b}(\theta)  \ .
    \label{Izzzmodel1}
\end{equation}
Here, $F^{zzz}_{a,b}(\theta)=h_a(\theta)h^{r,z}(\theta)h^{R,zz}(\theta)$ and $G^{zzz}_{a,b}(\theta)=g_a(\theta)h^{r,z}(\theta)h^{R,zz}(\theta)$. The exact expression of $F^{zzz}_{a,b}(\theta)$ and $G^{zzz}_{a,b}(\theta)$ can be found in Appendix B, Eq.~(\ref{F_abzzzapp}) and Eq.~(\ref{G_abzzzapp}), respectively.

In the $zxx$ direction, the shift vector and the density of state scale similarly to the $zzz$ direction. However, for the transition intensity, we have $|r_{cv}^{x}|^2=\epsilon^{-\frac{2}{a}}h^{r,x}_{a}(\theta)$. By multiplying all these three quantities, $I^{zxx}(\epsilon,\theta)$ is found as
\begin{equation}
    I^{zxx}(\epsilon,\theta)=\frac{k_0}{\epsilon^{2-b+\frac{1}{a}}}F^{zxx}_{a,b}(\theta)+\frac{1}{\epsilon^{2-b}}G^{zxx}_{a,b}(\theta) \ .
    \label{Izxxmodel1}
\end{equation}
Here, $F^{zxx}_{a,b}(\theta)=h_a(\theta)h^{r,x}_a(\theta)h^{R,zx}(\theta)$ and $G^{zxx}_{a,b}(\theta)=g_a(\theta)h^{r,x}_a(\theta)h^{R,zx}(\theta)$. The exact expression of $F^{zxx}_{a,b}$ and $G^{zxx}_{a,b}$ can be found in Appendix B, Eq.~(\ref{Fzxxapp}) and Eq.~(\ref{Gzxxapp}), respectively.

For Model 2, the evaluation of the shift conductivity would be similar to the case of Model 1 in the $zzz$ direction. The shift vector scales as in  Model 1, $R_{cv,z}^z=\epsilon^{-2+b}\tilde{h}^{R,zz}_{a,b}(\theta)$ and the transition intensity scales as $|r_{cv}^z|^2=\epsilon^{-2}\tilde{h}^{r,z}(\theta)$. In Model 2, the Jacobian of the variable change would be $|\det \tilde{J}|=\tilde{\rho}(\epsilon)\tilde{h}_a$, which is independent of $k_0$. By multiplying all these three quantities, we find that the $I^{zzz}$ tensor for Model 2 is given by 
\begin{equation}
\tilde{I}^{zzz}(\theta,\epsilon)=\frac{1}{\epsilon^{4-b-\frac{1}{a}}}\tilde{F}_{a,b}^{zzz}(\theta) \ .
\label{IzzzModel2}
\end{equation}
Here, $\tilde{F}_{a,b}^{zzz}(\theta)=\tilde{h}^{r,z}(\theta)\tilde{h}^{R,zz}_{a,b}(\theta)\tilde{h}_a$ (see Appendix B, Eq.~(\ref{tildeF_abzzzapp}) for details). Note that there is no contribution from $k_0$ to the shift conductivity in $zzz$ direction for Model 2, and the reason is that the joint density of state Eq.~(\ref{density2}) is independent of $k_0$ in this model.

In the $zxx$ direction, however, the result depends on $k_0^2$ as shown in Table~\ref{tab:a2b2}. The $k_0$ dependence in this direction comes from the transition intensity $|r_{cv}^x|^2=k_0^2 \epsilon^{-2/a}h^{r,x}_a(\theta)+\epsilon^{-1/a}g^{r,x}_a(\theta)$. Here, $h^{r,x}_a(\theta)$ and $g^{r,x}_a(\theta)$ are the angular parts of the transition intensity. The other two factors, shift vector and joint density of state, remain the same as the $zzz$ direction. Finally, by considering these three terms, we find that $I^{zxx}(\theta,\epsilon)$ can be written as
\begin{equation}
\tilde{I}^{zxx}(\theta,\epsilon)=\frac{k_0^2}{\epsilon^{2-b+1/a}}\tilde{F}^{zxx}_{a,b}(\theta)+\frac{1}{\epsilon^{2-b}}\tilde{G}_{a,b}^{zxx}(\theta) \ .
\end{equation}
Here, $\tilde{F}_{ab}^{zxx}$ and $\tilde{G}_{ab}^{zxx}$ are angular dependent function similar to $\tilde{F}^{zzz}_{ab}$.

Now we are ready to discuss the overall scaling behavior of the shift conductivity for Model 1 and Model 2.
Considering the frequency dependence of all the factors discussed above, one can see that it depends strongly on the choice of $a$ and $b$ exponent. Let us first investigate the effect of the $b$ coefficient in the shift conductivity. As can be seen from Eq.~(\ref{Izzzmodel1}), in order to find the most singular behavior for shift conductivity, it would be convenient to fix the $b$ exponent as the minimum integer as possible. However, for $b=1$, the Hamiltonian does not break the inversion symmetry and the shift conductivity would vanish (one can also see that all the angular functions would vanish explicitly for the choice of $b=1$ as seen in Appendix B). The next relevant choice is $b=2$, which would break the inversion symmetry of the Hamiltonian, leading to a finite shift conductivity. Hence we use $b=2$ for the following discussions even though the exponent $b$ is explicitly shown in all the formulas.

For Model 1, when $a$ is an even integer, we find that the second term of Eq.~(\ref{Izzzmodel1}) vanishes after integrating over $\theta$, and only the first term contributes to the shift conductivity. In this case, we can see that the shift conductivity scales as
\begin{equation}
\sigma^{zzz}_{\rm shift}(0;\omega,-\omega) \sim k_0\omega^{-4+b+\frac{1}{a}} \ \ \ (a = {\rm even}) \,.
\label{shiftscalingzzzevena}
\end{equation}
If $a$ is an odd integer, then the first term in Eq.~(\ref{Izzzmodel1}) vanishes, and only the second term can contribute, leading to  
\begin{equation}
\sigma^{zzz}_{\rm shift}(0;\omega,-\omega) \sim\omega^{-4+b+\frac{2}{a}} \ \ \ (a = {\rm odd}) \,.
\label{shiftscalingzzzodda}
\end{equation}

Similar power-law behaviors can be found for the $zxx$ and $xzx$ directions as well. Let us now consider the $zxx$ direction for simplicity, but one can generalize the following argument to the case of the $xzx$ direction. Similar to the $zzz$ direction, when we integrate over $\theta$ for an even integer $a$, only the first term in Eq.~(\ref{Izxxmodel1}) survives and the $G^{zxx}_{a,b}(\theta)$ would vanish. However, when $a$ is odd, the second term survives and the $F_{a,b}^{zxx}(\theta)$ vanishes. Thus, we find that the scaling behavior of the shift conductivity in $zxx$ directions is as follows when $a$ is even
\begin{equation}
    \sigma_{\rm shift}^{zxx}(0;\omega,-\omega) \sim k_0 \omega^{-2+b-1/a} \ \ \ (a = {\rm even}) \,,
    \label{shiftscalingzxxevena}
\end{equation}
and when $a$ is odd, we have
\begin{equation}
    \sigma_{\rm shift}^{zxx}(0;\omega,-\omega) \sim \omega^{-2+b} \ \ \ (a = {\rm odd}) \, .
    \label{shiftscalingzxxodda}
\end{equation}

For Model 2, the function $\tilde{F}_{a,b}^{zzz}(\theta)$ and $\tilde{F}^{zxx}_{a,b}(\theta)$ vanish after evaluating the integral over the angular part when $a$ is an odd integer. Hence, for odd integer $a$, the shift conductivity for the $zzz$ direction vanishes and it is finite only for even integer $a$. 
For the $zxx$ direction, only $\tilde{G}^{zxx}_{a,b}(\theta)$ part contributes for odd integer $a$ while both terms can contribute in the case of even integer $a$.
Therefore, when $a$ is even, we can write the scaling of the shift conductivity for Model 2 as follows. 
\begin{equation}
    \tilde{\sigma}^{zzz}_{\rm shift}(0;\omega,-\omega)\sim \omega^{-4+b+\frac{1}{a}} \ \ \ (a = {\rm even}) \,,
    \label{shiftscalingzzzevena2}
\end{equation}
\begin{equation}
   \tilde{\sigma}^{zxx}_{\rm shift}(0;\omega,-\omega)\sim k_0^2\omega^{-2+b-\frac{1}{a}} \ \ \ (a = {\rm even}) \, .
   \label{shiftscalingzxxevena2}
\end{equation}

When $a$ is odd, we get
\begin{equation}
   \tilde{\sigma}^{zxx}_{\rm shift}(0;\omega,-\omega)\sim \omega^{-2+b}  \ \ \ (a = {\rm odd}) \, . 
   \label{shiftscalingzxxodda2}
\end{equation}

As we discussed earlier, we may focus on the choice of $b=2$ in these models. Now let us consider both cases of $a=2,\,b=2$ and $a=1,\,b=2$. For Model 1 and Model 2 with $a=2,\,b=2$, in the $zzz$ direction, by using Eq.~(\ref{shiftscalingzzzevena}) and Eq.~(\ref{shiftscalingzzzevena2}), we obtain $\sigma_{\rm shift}^{zzz}\sim k_0/\omega^{3/2}$ and $\tilde{\sigma}_{\rm shift}^{zzz}\sim 1/\omega^{3/2}$, respectively. In comparison to other systems with singular behavior in the shift conductivity, such as Weyl semimetals, our models show even more singular behavior. In Type I and Type II Weyl semimetals, $\sigma_{\rm Weyl, I}^{zzz} \sim \omega^{0}$ and $\sigma^{zzz}_{\rm Weyl, II}\sim 1/\omega$ \cite{shiftweyl}, respectively. The exact expression of the shift conductivity of Model 1 and Model 2 in the case of $a=2\,,b=2$ can be found in Table~\ref{tab:a2b2}. For the second case, $a=1,\,b=2$, we can see that the shift conductivity is constant in both models as shown in Table~\ref{tab:a1b2}.

\subsection{Second harmonic generation}

The SHG in these classes of Hamiltonian is the sum of all four quantities Eq.~(\ref{SHG1})-Eq.~(\ref{SHG4}), which we are going to investigate in detail. For two-band systems, we can simplify the expression of the SHG response as follows.
\begin{align}
\label{SHG2pItwoband}
    &\sigma^{ijm}_{2p,I}(\omega)=\frac{i \pi e^3}{2\hbar^2 \omega^2}\int [dk] M_{2pI,cv}^{ijm}(k)f_{vc}\delta(2\omega-\omega_{cv}),\\&
    \label{SHG2pIItwoband}
    \sigma^{ijm}_{2p,II}(\omega)=\frac{i \pi e^3}{2\hbar^2 \omega^2}\int [dk] M_{2pII,cv}^{ijm}(k)f_{vc}\delta(2\omega-\omega_{cv}),\\&
        \label{SHG1pItwoband}
    \sigma^{ijm}_{1p,I}(\omega)=\frac{i \pi e^3}{2\hbar^2 \omega^2}\int [dk] M_{1pI,cv}^{ijm}(k)f_{vc}\delta(2\omega-\omega_{cv}),\\&
        \label{SHG1pIItwoband}
    \sigma^{ijm}_{1p,II}(\omega)=\frac{i \pi e^3}{2\hbar^2 \omega^2}\int [dk] M_{1pII,cv}^{ijm}(k)f_{vc}\delta(2\omega-\omega_{cv}),
\end{align}
where 
\begin{align}
\label{M2pI}
    &M_{2pI,cv}^{ijm}=v^{i}_{vc}w_{cv}^{jm},\\&
    \label{M2pII}
    M_{2pII,cv}^{ijm}=\frac{-4v_{vc}^{i}\big[v_{cv}^{j},v_{cc}^{m}\big]_{+}}{\omega_{cv}},\\&
        \label{M1pI}
        M_{1pI,cv}^{ijm}=w^{ij}_{vc}v_{cv}^{m}+w_{vc}^{im}v^{j}_{cv},\\&
        \label{M1pII}
     M_{1pII,cv}^{ijm}=\frac{2v_{vc}^{i}}{\omega_{cv}}\big[v_{cc}^{j},v_{cv}^{m}\big]_{+}-\frac{v_{cc}^{i}}{\omega_{cv}}\big[v_{cv}^{j},v_{vc}^{m}\big]_{+},
\end{align}
where $\big[v_{nm}^i,v_{mp}^j\big]_{+}=v_{nm}^{i}v_{mp}^j+v_{nm}^{j}v_{mp}^i$, and $n$, $m$ and $p$ are the band index.
Note that we used $v^i_{cc}=-v^i_{vv}$, which holds in our models. 
Note that 
the imaginary part of $M^{ijm}$ gives rise to the real part of the SHG.

Let us first consider the $zzz$ direction to investigate the effect of each term on the real part of the SHG. In the $zzz$ direction, by using the fact that $v^{i}_{cv}=v^{i*}_{vc}$, it is easy to see that Eq.~(\ref{M2pII}) and Eq.~(\ref{M1pII}) are completely real and do not contribute to the real part of the SHG. In the $zzz$ direction, $M_{2p,I}^{zzz}=-\frac{1}{2}M_{1p,I}^{zzz}$, which is proportional to $|r_{cv}^z|^2R_{cv,z}^z$ (see Appendix D Eq.~(\ref{Msrelation})). Thus, remarkably, we can write the SHG response in $zzz$ direction in terms of the shift conductivity in the $zzz$ direction.
\begin{align}
\label{SHGzzztoShift}
    &\operatorname{Re}\big[\sigma_{SHG}^{zzz}(2\omega;\omega,\omega)\big]=\\\nonumber
    &\sigma_{\rm shift}^{zzz}(0;2\omega,-2\omega)-\frac{1}{2}\sigma_{\rm shift}^{zzz}(0;\omega,-\omega)\,.
\end{align}

Now we investigate the SHG for the $zxx$ direction. In our choice of Hamiltonians, considering the $zxx$ direction, $M_{1pI}^{zxx}$ vanishes because there is no crossing terms such as $k_xk_z$ in the Hamiltonian. Thus, the finite contributions come from $M^{zxx}_{2pI}$, $M^{zxx}_{2pII}$ and $M^{zxx}_{1pII}$. By some calculation (see Appendix E), we can show that these quantities can be written in terms of the shift conductivity. As a result, the SHG in the $zxx$ direction can be written in terms of the shift conductivity as follows.
\begin{align}
\label{SHGzxxtoshift}
&\operatorname{Re}\big[\sigma^{zxx}_{SHG}(2\omega;\omega,\omega)\big] = \nonumber \\
&-3\sigma^{zxx}_{\rm shift}(0;2\omega,-2\omega)+2\sigma^{xzx}_{\rm shift}(0;2\omega,-2\omega) 
+\frac{1}{2}\sigma^{zxx}_{\rm shift}(0;\omega,-\omega)\,.
\end{align}
Recall that the shift conductivity is related to the shift vector, which is a quantum geometric quantity in momentum space and has topological information about the wavefunction. Hence, interestingly, the real part of the SHG in both models (Model 1 and Model 2) can also be expressed in terms of such quantum geometric quantities. (See appendix D)

\section{Summary and Discussion}

In the present work, we show that the NLSM in the presence of an external {\it dc} electric field exhibits a large SHG response at finite doping and in the low frequency limit. 
It is given by $\sigma_{SHG}(2\omega;\omega,\omega)\sim k_0/\mu^2$ when $vk_0>\mu$, which is a more singular response than that of Weyl semimetals. 
On the other hand, in the regime $vk_0\ll \mu$, we obtain 
$\sigma_{SHG}(2\omega,\omega,\omega)\sim k_0/\mu^2+1/\mu$. Note that, in the limit $k_0\rightarrow0$, the result reduces to the case of Weyl semimetals in an external {\it dc} electric field, $\sigma_{SHG}(2\omega,\omega,\omega)\sim 1/\mu$. 

\widetext

\begin{table}[]
    \begin{tabular}{ |p{3cm}|p{3cm}|p{3cm}|  }

    \hline
    \multicolumn{3}{|c|}{Choice of parameters $a=2$ and $b=2$} \\
    \hline
    Shift conductivity & Model 1 & Model 2 \\
    \hline

    $\sigma_{\text{shift}}^{zzz}(0;\omega,-\omega)$ & 
    $\frac{e^3}{\hbar^2}\frac{2k_0\sqrt{\frac{2\pi}{t_x}}t_z \Gamma(3/4)^2}{\omega^{3/2}}$ & $\frac{e^3}{\hbar^2}\frac{\sqrt{\frac{2\pi}{\tilde{t}_x}}tz \Gamma(3/4)^2}{\omega^{3/2}}$ \\ 
    $\sigma_{\text{shift}}^{zxx}(0;\omega,-\omega)$ & $\frac{e^3}{\hbar^2}\frac{4\pi k_0\sqrt{2t_x}t_z K(1/2)}{21v_y^2\omega^{1/2}}$ & $\frac{e^3}{\hbar^2}\frac{8\pi k_0^2\sqrt{2\tilde{t}
    _x}t_z K(1/2)}{21v_y^2\omega^{1/2}}$ \\
    $\sigma_{\text{shift}}^{xzx}(0;\omega,-\omega)$ & $\frac{e^3}{\hbar^2}\frac{5\pi k_0\sqrt{2t_x}t_z K(1/2)}{21v_y^2\omega^{1/2}}$ & $\frac{e^3}{\hbar^2}\frac{10\pi k_0^2\sqrt{2\tilde{t_x}}t_z K(1/2)}{21v_y^2\omega^{1/2}}$ \\
    \hline
    \end{tabular}
    \caption{The shift current conductivity for the case of $a=2$ and $b=2$ for both Model 1 and Model 2. For this parameter choice, we can see divergent behavior in both models along $zzz$, $zxx$, and $xzx$ directions. The $zzz$ conductivity has more singular behavior than that of other directions for both models. Note that in $zzz$ direction, in the Model 2, the shift conductivity is independent of $k_0$. Here $\Gamma(x)$ is the Euler gamma function and $K(x)$ the complete elliptic integral of the first kind.}
    \label{tab:a2b2}
\end{table}

\begin{table}[]
    \centering
    \begin{tabular}{ |p{3cm}|p{3cm}|p{3cm}|  }
    \hline
    \multicolumn{3}{|c|}{Choice of parameters $a=1$ and $b=2$} \\
    \hline
    Shift conductivity & Model 1 & Model 2 \\
    \hline

    $\sigma_{\text{shift}}^{zzz}(0;\omega,-\omega)$ & 
     $\frac{e^3}{\hbar^2}\frac{\pi^2tz}{t_x^2}$ & $0$ \\ 
    $\sigma_{\text{shift}}^{zxx}(0;\omega,-\omega)$ & $\frac{e^3}{\hbar^2}\frac{\pi^2 t_z}{16 v_y^2}$ & $\frac{e^3}{\hbar^2}\frac{\pi^2 t_z}{8v_y^2}$ \\
    $\sigma_{\text{shift}}^{xzx}(0;\omega,-\omega)$ & $\frac{e^3}{\hbar^2}\frac{3\pi^2 t_z}{32 v_y^2}$ & $\frac{e^3}{\hbar^2}\frac{3\pi^2 t_z}{16 v_y^2}$ \\
    \hline
    \end{tabular}
\caption{The shift conductivity for the case of $a=1$ and $b=2$ for both Model 1 and Model 2. The conductivities in this case are constant in both models and independent of $k_0$.}
    \label{tab:a1b2} 
\end{table}

\endwidetext

We then consider the shift conductivity and SHG in two models of the NLSM, Model 1 and Model 2, with intrinsically broken inversion symmetry. 
The shift conductivity in both models strongly depends on the dispersion relation of the NLSM.
The results are shown in Table~\ref{tab:a1b2} and Table~\ref{tab:a2b2}.
We also show that the real part of the SHG is related to the shift current as shown in Eq.~\ref{SHGzxxtoshift} and Eq.~\ref{SHGzzztoShift}.
To estimate an order of magnitude of the results in Table~\ref{tab:a2b2} let us consider the system in the low temperature and low frequency limit (Terahertz regime). Using $\omega \sim 1 \ {\rm THz}$, we can see that $\sigma_{\text{w}}=\frac{e^3}{\hbar^2\omega}\sim 0.01 A/V^2$ where $\sigma_{\text{w}}$ is the conductivity of the Type II Weyl semimetal\cite{shiftweyl}. However, if we choose the exponent $a=2$ and $b=2$ in the Model 1 for the NLSM and consider $t_x \sim t_z$ and $v_y \sim t_x $, we find that $\sigma^{zzz}_{\text{NLSM}}\sim \sigma_{\text{w}} \sqrt{(\frac{t_xk_0^2}{\hbar\omega})}$. In the low frequency limit, if we choose
$\hbar \omega \ll t_x k_0^2 \sim 100 \hbar\omega$, we get $\sigma^{zzz}_{\text{NLSM}}\sim 0.1 A/V^2$. Hence the response of the NLSM can be an order of magnitude larger than the response of the Type II Weyl semimetal. 


\begin{acknowledgments}
This work was supported by the Natural Sciences and Engineering Research Council of Canada (NSERC) and the Centre for Quantum Materials at the University of Toronto.
\end{acknowledgments}

\bibliography{bibliography}

\appendix
\widetext

\section{Second order response}
In the systems with multi-band structure there exhibit two kind of polarization due to the interband  and intraband interactions, which we will denote them as ${\bf P}_{e}$ and ${\bf P}_{i}$ respectively. 

\begin{equation}
    {\bf \hat{P}}={\bf \hat{P}}_{e}+{\bf \hat{P}}_{i}\,,
\end{equation}
where 
\begin{equation}
   {\bf \hat{P}}_{e}=e \int [dk]\sum_{nm}{\bf r}_{nm}a^{\dagger}_{n}(k)a_{m}(k)\,,
   \label{Peapp}
\end{equation}
\begin{equation}
    {\bf \hat{P}}_{i}=e \int [dk] \sum_{n}(a^{\dagger}_{n}(k)\partial_{\bf k} a_{n}(k)-i\bm{\xi}_{nn}a^{\dagger}_{n}(k)a_{n}(k))\,,
\end{equation}
and 
\begin{align}
    \bm {\xi}_{nm}=i u_n(k)^{*}\partial_{{\bf k}}u_m(k)\\\nonumber 
    {\bf r}_{nm}=\bm{\xi}_{nm}\,\,\,\,\,\text{when}\,\,\,\,n\neq m\,.
\end{align}
Here, ${\bf r}_{nm}$ is the off-diagonal element of the position operator and the $\bm{\xi}_{nn}$ can be considered as the berry connection of the $n$th band.

By applying an optical field on the system we can find the linear and non-linear susceptibility $\big<{\bf P}(\omega_\Sigma)\big>=\chi_{0}+\chi_{1}E(\omega_{\Sigma})+\chi_{2}(\omega_\Sigma,\omega_1,\omega_2)E_{\omega_1}E_{\omega_2}+...$
where $\omega_{\Sigma}=\omega_1+\omega_2$.
To find the susceptibility, we need to find the generalized distribution function in the presence of the applied electric field. We define generalized distribution function as $\rho_{nm}=<a^{\dagger}_{n}(k)a_{m}(k)>$ which satisfy the following equation  
\begin{equation}
    \frac{\partial \rho_{nm}}{\partial t}+i\omega_{mn}\rho_{mn}=\frac{e}{i\hbar}\sum_{lb}E^{b}(\rho_{ml}r^{b}_{ln}-r^{b}_{ml}\rho_{ln})-\frac{e}{\hbar}E^{b}\rho_{mn;b}\,,
    \label{roHeq}
\end{equation}
where the the generalized derivative be defined as $\rho_{mn;b}=\partial_{k_b}\rho_{nm}-i(\xi_{nn}^b-\xi_{mm}^b)\rho_{nm}$ and $\omega_{nm}=\omega_n-\omega_m$. 

By solving the Eq.~(\ref{roHeq}) perturbatively on the electric field, we can find the non-linear contributions. let us consider the following form for the density distribution function 
\begin{equation}
    \rho_{nm}=f_n \delta_{nm}+\tilde{\rho^{(1)}}+\rho^{(2)}\,,
\end{equation}
where $f_{n}$ is the Fermi distribution. The $\tilde{\rho^{(1)}}$ shows the first-order contribution, and $\rho^{(2)}$ shows the second-order contribution in the electric field. 

By solving Eq.~(\ref{roHeq}) to the first order in electric field we can find the following solution
\begin{align}
    &\tilde{\rho}^{(1)}_{nm}=\rho_{FS}^{(1)}+\rho^{(1)}\,,\\
    &\rho_{FS}^{(1)}=-\delta_{nm}\frac{ie}{\hbar}\sum_{b\beta}\frac{1}{\omega_\beta}\frac{\partial f_n}{\partial k_b}E_\beta^b e^{-i\omega_\beta t}\,,\\
    &\rho^{(1)}=\frac{e}{\hbar}\sum_{b\beta} \frac{r^{b}_{nm}f_{nm}}{\omega_{mn}-\omega_\beta}E_\beta^{b}e^{-i\omega_\beta t} \,.
    \label{rho1solapp}
\end{align}
The $\rho_{FS}^{(1)}$ denotes the first-order contribution coming from the Fermi surface of the system at low temperature (because of the derivative of the distribution function). Let us assume there is no Fermi surface in the system, thus we can ignore the first contribution and at the second-order perturbation we can find the following solution
\begin{equation}
    \rho^{(2)}=\frac{ie}{\hbar(\omega_{nm}-\omega_\Sigma)}[\rho^{(1)}_{mn;c}+i\sum_{l}(\rho_{ml}^{(1)}r_{ln}-r_{ml}\rho_{ln}^{(1)})]\,.
    \label{rho2solapp}
\end{equation}

Finally by using Eq.~(\ref{rho2solapp}) in Eq.~(\ref{Peapp}) we can find the susceptibilities for the interband  contribution as following
\begin{align}
    \label{suse}
    \frac{\chi_e^{(2),abc}(\omega_{\Sigma};\omega_{\beta},\omega_{\alpha})}{C}=&i\sum_{nmk}\frac{r^{a}_{nm}f_{nm}}{\omega_{nm}-\omega_\Sigma}\bigg(\frac{r^b_{mn}}{\omega_{mn}-\omega_{\beta}}\bigg)_{;c}-\\\nonumber&
    \sum_{nlmk}\frac{r^{a}_{nm}}{\omega_{nm}-\omega_\Sigma}\bigg(\frac{r^{b}_{ml}r^{c}_{ln}f_{lm}}{\omega_{ml}-\omega_{\beta}}-\frac{r^{c}_{ml}r^{b}_{ln}f_{nl}}{\omega_{ln}-\omega_\beta}\bigg)\,.
\end{align}
Here, $C=e^3/\hbar^2V$. 

To calculate the intraband  contribution it would be convenient to start with the intraband  current  ${\bf J}_i={\bf J}-{\bf J}_e$, where ${\bf J}$ is the total current and ${\bf J}_e$ is the interband   current. We can find the total current as following 
\begin{equation}
    {\bf J}= \frac{e}{V}\sum_{nmk}{\bf v}_{nm}a_{n}^{\dagger}(k)a_{m}(k)\,,
    \label{Jtotapp}
\end{equation}
where $V$ is the volume of the system and $v^a_{nm}=\big<n|\partial_{k_a}H|m\big>=\omega_{n;a}\delta_{nm}-i\omega_{mn}r^a_{nm}$. And for the interband   current we have the following expressions 
\begin{equation}
    {\bf J}_{e}=\frac{-i}{\hbar}[{\bf P}_{e},H_0]+\frac{i}{\hbar}[{\bf P}_{e},{\bf P}_{i}+{\bf P}_{e}]\,.
\end{equation}
The contribution of the first term can be written as following
\begin{equation}
    \frac{-i}{\hbar}[{\bf P}_{e},H_0]=\frac{e}{V}\sum_{nmk}\omega_{mn}{\bf r}_{nm}a_{n}^{\dagger}(k)a_{m}(k)\,.
\end{equation}
And the contribution of the second term can be written as following
\begin{equation}
    [P^a_{e},P^b_{i}+P^b_{e}]=\frac{-ie^2}{V^2}\sum_{nmk}\big(r^a_{nm;b}+i\sum_{p}(r_{np}^ar_{pm}^b-r_{np}^br_{pm}^a)\big)a^{\dagger}_{n}(k)a_{m}(k)\,.
\end{equation}

Finally, by using $\frac{d{\bf P}_i}{dt}={\bf J}-{\bf J}_{e}$ we can find the susceptibility for the intraband  contribution  
\begin{align}
    \frac{\chi^{abc}_{i}(\omega_{\Sigma};\omega_{\beta},\omega_{\alpha})}{C}= \frac{i}{\omega_{\Sigma}^2}\sum_{nmk}\frac{\omega_{nm;a}r_{nm}^br_{mn}^cf_{mn}}{\omega_{nm}-\omega_{\beta}}+\frac{1}{i\omega_{\Sigma}}\sum_{nmk}\frac{r_{nm;a}^cr_{mn}^bf_{nm}}{\omega_{mn}-\omega_\beta}\,.
    \label{susi}
\end{align}

\subsection{Shift and injection}
By looking at Eq.~(\ref{suse}) and Eq.~(\ref{susi}) we can see in the dc limit ($\omega_{\Sigma}\rightarrow 0$) the dominant contribution comes from $\chi_{i}$, so we can neglect the $\chi_e$ contribution in the dc limit. We can find the shift current and the injection current just by investigation the divergence behavior of Eq.~(\ref{susi})(the first term is injection and the second term is the shift current), however it would be more intuitively useful if we look at the intraband currents in the presence of the electric field (which is the dominant contribution). 
\begin{equation}
    J^{a}_i=\frac{e}{V}\sum_{nmk} [\omega_{n;a}\delta_{nm}-\frac{e}{\hbar}({\bf E\times\Omega}_{n})^{a}\delta_{nm}-\frac{e}{\hbar}{\bf E.r}_{nm;a}]a^{\dagger}_{n}(k)a_{m}(k)\,,
\end{equation}
\begin{equation}
    J^{a}_i=\frac{e}{V}\sum_{nmk} [\omega_{n;a}\delta_{nm}-\frac{e}{\hbar}({\bf E\times\Omega}_{n})^{a}\delta_{nm}-\frac{e}{\hbar}{\bf E.r}_{nm;a}](f_{n}\delta_{nm}+\rho_{FS}^{(1)}+\rho_{nm}^{(1)}+\rho_{nm}^{(2)}+\rho_{FS}^{(2)})\,.
\end{equation}
Now for the second order response we have 
\begin{align}
&J_{i,I}^a=\sum_{nmk}\omega_{n;a}\delta_{nm}\rho_{nm}^{(2)}\,,\\&
J_{i,II}^a=-\sum_{nmk}\frac{e}{\hbar}{\bf E.r}_{nm;a}\rho_{nm}^{(1)}\,,\\&
J_{i,FS_I}^a=\sum_{nmk}\omega_{n;a}\delta_{nm}\rho_{FS}^{(2)}\,,\\&
J_{i,FS_{II}}^a=-\sum_{nmk}\frac{e}{\hbar}({\bf E\times \Omega}_n)^a\delta_{nm}\rho_{FS}^{(1)}\,.
\end{align}
The real part of the $J_{i,I}^a$ and $J_{i,II}^a$ are responsible for the injection and shift current respectively. Two scenarios can lead to BPVE. In the first scenario the charge carriers relax momentum asymmetrically into $\pm k$ direction via collisions with electrons, phonons or impurities which makes a net current. The second scenario the light-matter interaction give rise to the BPVE. In the injection current processes, light pumps carriers into the conduction band asymmetrically at $\pm k$ in the BZ and this leads to the net current. To explain the injection current in the wave packet approximation we can consider an electron with velocity $v^{a}_n$ and we can define the current as $J^a=\frac{e}{V}\sum_{nk}f_n v_n^a$. By applying an optical field to the system, current carrying states are injected to the conduction band. 
\begin{equation}
    \frac{d J_{inj}}{dt}=\frac{e}{V}\sum_{nk}\frac{df_n}{dt}v_n^a\,,
\end{equation}
where $\frac{df_n}{dt}$ is given by fermi golden rule
\begin{equation}
    \frac{df_c}{dt}=\frac{2\pi e^2}{\hbar^2}\sum_{v}|E(\omega).r_{cv}|^2\delta(\omega_{cv}-\omega)\,,
\end{equation}
and breaking inversion symmetry $ \frac{d}{dt}f_c(k)\neq  \frac{d}{dt}f_c(-k)$ allow us to have a net current. $c$ and $v$ index related to conduction and valenced band respectively. Note that in a system with a time-reversal symmetry (such as Model 1 Eq.~(\ref{NLSM1}) and Model 2 Eq.~(\ref{NLSM2})) the injection current vanishes because it is an odd function under time-reversal symmetry and will change sign under transformation $\omega_{\Sigma}\rightarrow -\omega_{\Sigma}$.

In the shift current process, Inversion symmetry breaking along with the separation of the center of the charge of the valence and conduction band give rise to the net current. The separation of center of charge make the dipole velocity oscillation and the inversion breaking make the net current to be non-zero.   

\section{Shift current in Model 1 and Model 2}
For the two-band systems, where the induced light is polarized in the $b$ direction, the shift current in Eq.~(\ref{susi}) can be written as following (see Appendix D)
\begin{equation}
    \sigma_{sh}^{abb}(0;\omega,-\omega)=\frac{\pi e^3}{\hbar^2}\sum_{cv}\int_{BZ}[dk]I^{abb}_{cv}(k)f_{cv}\delta(\omega_{cv}-\omega)\,,
\end{equation}
where 
\begin{equation}
    I^{abb}_{cv}(k)=-\sum_{ijm}\frac{1}{4d^3}(d_m d_{i,b}d_{j,ab}-d_md_{i,b}d_{j,a}\frac{d_{,b}}{d})\epsilon_{ijm}\,.
\end{equation}
Here $d_{i}(k)$ are coefficients of the Pauli matrices in the Hamiltonian Eq.~(\ref{HappD}).

In the following subsections we are going to calculate the $I^{abb}$ tensor for both Model 1 and Model 2 in the $zzz$ and $zxx$ direction. 

\subsection{Model 1}

Let us only look at the $I^{zxx}$ component for simplicity. 
\begin{equation}
    I^{zxx}=t_x^3t_zv_ya(-1+b)\frac{k_x^2}{4(k_x^2+k_y^2)}\frac{(\sqrt{k_x^2+k_y^2}-k_0)^{-2+3a}k_z^b}{\big(t_x^2(\sqrt{k_x^2+k_y^2}-k_0)^{2a}+t_z^2k_z^{2b}+v_y^2k_z^2\big)^{5/2}}\,.
    \label{Izxxapp2}
\end{equation}
As we discussed in section IV.A the $b$ exponent can break the symmetry if it has chosen to be an even integer. Also, in the Eq.~(\ref{Izxxapp2}), we can see the integration of the function over the Brillouin zone is nonvanishing only for even choice of $b$ exponent. To proceed we are going to use an approximation in which $\epsilon\sim \sqrt{\sqrt{k_x^2+k_y^2}-k_0)^{2a}+v_y^2k_z^2}$. As long as $b>1$ we can neglect $t_z k_z^b$ term in compare with $v_y k_z$ in the low frequency limit. By this approximation we can use the following variable change
\begin{align}
\label{varchange}
 &k_z= \frac{1}{v_y}d \cos(\theta)\,,\\\nonumber&
 \sqrt{k_x^2+k_y^2}=\pm (\frac{1}{t_x}d \sin(\theta))^{1/a}+k_0\,,\\\nonumber&
 \theta \in (0,\pi) \,\,\,\, \text{for} \,\,\,\,\ \pm\,,
\end{align}
where the determinant of the Jacobi matrix would be 
\begin{equation}
    \det|J_\pm|=\frac{\csc(\theta)(\frac{d \sin(\theta)}{t_x})^{1/a}(k_0\pm(\frac{d \sin(\theta)}{t_x})^{1/a})}{a v_y}\,.
    \label{Jacobian}
\end{equation}
Let us first look at the $+$ sign. Using the variable change Eq.~(\ref{varchange}) in Eq.~(\ref{Izxxapp2}) in the low frequency limit we can find 
\begin{equation}
    I_{+}^{zxx} det|J_+|=t_x^3t_zv_ya(-1+b)\csc(\theta)\cos(\phi)^2\frac{(\frac{d \sin(\theta)}{t_x})^{\frac{-1}{a}+3}(\frac{d \cos(\theta)}{v_y})^b(k_0+(\frac{d \sin(\theta)}{t_x})^{1/a})}{d^5}\,.
\end{equation}
Also we can do the same with $-$ sign in which we find the following expression 
\begin{equation}
    I_{-}^{zxx} det|J_-|=t_x^3t_zv_ya(-1+b)\csc(\theta)\cos(\phi)^2\frac{(-1)^{3a}(\frac{d \sin(\theta)}{t_x})^{\frac{-1}{a}+3}(\frac{d \cos(\theta)}{v_y})^b(k_0-(\frac{d \sin(\theta)}{t_x})^{1/a})}{d^5}\,,
\end{equation}
\begin{equation}
    \sigma^{zxx}_{shift}(0;\omega,-\omega)=\int_{0}^{\infty}\int_{0}^{\pi}I_+^{zxx}det|J_+|\delta(2d-\omega)d\theta dd+\int_{0}^{\infty}\int_{0}^{\pi}I_-^{zxx}det|J_-|\delta(2d-\omega)d\theta dd\,.
\end{equation}
Let us look at the structure of the $I_+$, after integrating over $\phi$, we find the following expression  
\begin{equation}
    I_{+}^{zxx}det|J_+|=\pi t_x^3t_zv_ya(-1+b)d^{-2+b-\frac{1}{a}}\big(k_0+(\frac{d \sin(\theta)}{t_x})^{1/a}\big)\bigg(\csc(\theta)\cos(\phi)^2(\frac{\sin(\theta)}{t_x})^{\frac{-1}{a}+3}(\frac{ \cos(\theta)}{v_y})^b\bigg)\,.
\end{equation}
The term proportional to $k_0$ can produce more singularity in the shift current (in the case that it's not zero because of the integral on $\theta$, for example it would be zero when $a=1$ and $b=2$). We can write the integral in the following way which make it easier to see the behavior of the shift current with respect to the energy $d$
\begin{equation}
     I_{+}^{zxx}det|J_+|=\frac{k_0}{d^{2-b+\frac{1}{a}}}f^{zxx}_{a,b}(\theta)+\frac{1}{d^{2-b}}g^{zxx}_{a,b}(\theta)\,,
\end{equation}
where 
\begin{equation}
    f^{zxx}_{a,b}(\theta)=3\pi t_zt_x^{1/a}
    v_ya(1-b)\bigg(\csc(\theta)\cos(\phi)^2(\sin(\theta))^{\frac{-1}{a}+3}(\frac{ \cos(\theta)}{v_y})^b\bigg)\,,
\end{equation}
and 
\begin{equation}
    g^{zxx}_{a,b}(\theta)=f^{zxx}_{a,b}(\theta)(\frac{\sin(\theta)}{t_x})^{1/a}\,.
\end{equation}
Now by considering $I^{zxx}_+\det|J_+| + I^{zxx}_-\det| J_-|$ we can find the $I^{zxx}(\theta,d)$ defined in Eq.~(\ref{Iabbshift}) as following 
\begin{equation}
    I^{zxx}(\theta,d)=\frac{k_0}{d^{2-b+\frac{1}{a}}}F^{zxx}_{a,b}(\theta)+\frac{1}{d^{2-b}}G^{zxx}_{a,b}(\theta)\,,
\end{equation}
where 
\begin{align}
    \label{Fzxxapp}
&F^{zxx}_{a,b}(\theta)= f^{zxx}_{a,b}(\theta)(1+(-1)^{a})\,,\\
    \label{Gzxxapp}
&G^{zxx}_{a,b}(\theta)= g^{zxx}_{a,b}(\theta)(1-(-1)^{a})\,.
\end{align}
Note that $F^{zxx}_{a,b}(\theta)$ is finite when $a$ exponent is an even number and $G^{zxx}_{a,b}(\theta)$ is finite only when $a$ exponent is an odd number.

We can do the same for $zzz$ direction. 
\begin{equation}
    I^{zzz}=t_xt_zv_y\frac{b(1-b)\big(\sqrt{k_x^2+k_y^2}-k_0\big)k_z^{-2+b}}{4\big(t_x^2(\sqrt{k_x^2+k_y^2}-k_0)^{2a}+t_z^2k_z^{2b}+v_y^2k_z^2\big)^{3/2}}\,.
\end{equation}

Using variable change in Eq.~(\ref{varchange})  we find the $I^{zzz}$ tensor can be written as follows 
\begin{equation}
    I_{+}^{zzz}det|J_+| =\frac{t_x^{-1/a}t_zv_y^{2-b}}{4a}b(-1+b)\cos(\theta)^{b-2}\sin(\theta)^{\frac{1}{a}}d^{-4+b+\frac{1}{a}}\big(k_0+d^{1/a}(\frac{\sin(\theta)}{t_x})^{1/a}\big)\,,
    \label{IJ+zzz}
\end{equation}
where by separating the angular part of the function we can simplify the above equation as following
\begin{equation}
    I_{+}^{zzz}det|J_+|=\frac{k_0}{d^{4-b-\frac{1}{a}}}f^{zzz}_{a,b}(\theta)+\frac{1}{d^{4-b-\frac{
    2}{a}}}g^{zzz}_{a,b}(\theta)\,.
\end{equation}
By considering $I^{zzz}(\theta,d)=I_{+}^{zzz}det|J_+|+I_{-}^{zzz}det|J_-|$ we can find 
\begin{equation}
     I^{zzz}(\theta,d)=\frac{k_0}{d^{4-b-\frac{1}{a}}}F^{zzz}_{a,b}(\theta)+\frac{1}{d^{4-b-\frac{
    2}{a}}}G^{zzz}_{a,b}(\theta)\,,
\end{equation}
where
\begin{align}
    &F_{a,b}^{zzz}(\theta)=\frac{t_x^{-1/a}t_zv_y^{2-b}}{4a}b(-1+b)\cos(\theta)^{b-2}\sin(\theta)^{1/a}k_0(1+(-1)^a)\,,
    \label{F_abzzzapp}\\
   & G_{a,b}^{zzz}(\theta)=\frac{t_x^{-2/a}t_zv_y^{2-b}}{4a}b(-1+b)\cos(\theta)^{b-2}\sin(\theta)^{2/a}(1-(-1)^a)\,.
    \label{G_abzzzapp}
\end{align}

\subsection{Model 2}
Model 2 Hamiltonian can be written as follows 
\begin{equation}
    \tilde{H}(k)= \tilde{t}_x(k_{x}^2+k_{y}^2-k_0^2)^a\tau_{x}+v_y k_z \tau_{y}+t_z k_{z}^b \tau_z \,,
\end{equation}
where the scaling of the $\tau_{x}$ and $\tau_y$ is different. 

The $I^{zxx}$ and $I^{zzz}$ can be written in the following way
\begin{equation}
    I^{zxx}=a^2(1-b)\tilde{t}_x^3t_zv_y\frac{k_x^2k_z^{b}(k_x^2+k_y^2-k_0^2)^{-2+3a}}{\big(\tilde{t}_x^2(k_x^2+k_y^2-k_0^2)^{2a}+v_y^2k_z^2+t_z^2k_z^{2b}\big)^{5/2}}\,.
    \label{Izxxmod2app}
\end{equation}
\begin{equation}
    I^{zzz}=b(1-b)\tilde{t}_xt_zv_y\frac{k_x^2k_z^{-2+b}(k_x^2+k_y^2-k_0^2)^{a}}{4\big(\tilde{t}_x^2(k_x^2+k_y^2-k_0^2)^{2a}+v_y^2k_z^2+t_z^2k_z^{2b}\big)^{3/2}}\,.
\end{equation}

In Model 2 we use the following transformation  
\begin{align}
\label{varchange2}
 &k_z= \frac{1}{v_y}d \cos(\theta)\,,\\\nonumber&
 k_x=\sqrt{\pm(\frac{d}{t_x}\sin(\theta))^{\frac{1}{a}}+k_0^2}\cos(\phi)\,,\\\nonumber&
 k_y=\sqrt{\pm(\frac{d}{t_x}\sin(\theta))^{\frac{1}{a}}+k_0^2}\cos(\phi)\,,\\\nonumber&
 \theta \in [0,\pi) \,\,\,\, \text{for} \,\,\,\,\ \pm\,,\\\nonumber&
 \phi \in [0,2\pi)\,.
\end{align}
The determinant of the Jacobian metrix for this transformation is given by the following expression 
\begin{equation}
    \det|J_{\pm}|=(\frac{d}{\tilde{t_x}})^{\frac{1}{a}}\frac{\sin(\theta)^{\frac{1}{a}-1}}{2av_y}\,.
\end{equation}

Using the variable change in Eq.~(\ref{Izxxmod2app}) the transformation we can find that 
\begin{equation}
    I^{zxx}_+det|J_+|=\frac{a(b-1)t_z\tilde{t}_x^{1/a}}{2v_y^b}d^{-2+b-\frac{1}{a}}\cos(\theta)^b\sin(\theta)^{2-1/a}\bigg(k_0^2+d^{1/a}(\frac{\sin(\theta)}{\tilde{t}_x})^{1/a}\bigg)\sin(\phi)^2\,.
\end{equation}
By considering $\tilde{I}^{zxx}(\theta,d)=I^{zxx}_+det|J_+|+I^{zxx}_-det|J_-|$ we find 
\begin{equation}
    \tilde{I}^{zxx}(\theta,d)=\frac{k_0^2}{d^{2-b+\frac{1}{a}}}\tilde{F^{zxx}_{ab}}+\frac{1}{d^{2-b}}\tilde{G}_{ab}^{zxx}\,,
\end{equation}
where 
\begin{align}
    &\tilde{F}_{ab}^{zxx}=\frac{\pi a(b-1)t_z\tilde{t}_x^{1/a}}{2v_y^b}\cos(\theta)^b\sin(\theta)^{2-1/a}k_0^2\big(1+(-1)^a\big)\\&
    \tilde{G}_{ab}^{zxx}=\frac{\pi a(b-1)t_z}{2v_y^b}\cos(\theta)^b\sin(\theta)^{2}\big(1-(-1)^a\big)\,.
\end{align}

We can do the same for $I^{zzz}$ in which we find  
\begin{equation}
    I^{zzz}_+det|J_+|=\frac{b(b-1)t_z\tilde{t}_x^{-1/a}}{8a v_y^{b-2}}d^{-4+b+\frac{1}{a}}\cos(\theta)^{b-2}\sin(\theta)^{1/a}\,,
\end{equation}
and by considering $\tilde{I}^{zzz}(\theta,d)=I^{zzz}_+det|J_+|+I^{zzz}_-det|J_-|$ we can find the final form of $I^{zzz}(\theta,d)$ as following 
\begin{equation}
    \tilde{I}^{zzz}(\theta,d)=\frac{1}{d^{4-b-\frac{1}{a}}}\tilde{F}_{ab}^{zzz}\,,
\end{equation}
\begin{equation}
    \tilde{F}^{zzz}_{a,b}(\theta)=\frac{b(b-1)t_z\tilde{t}_x^{-1/a}}{8a v_y^{b-2}}\cos(\theta)^{b-2}\sin(\theta)^{1/a}\big(1+(-1)^a\big)\,.
    \label{tildeF_abzzzapp}
\end{equation}

\section{Second Harmonic Generation in Nodal line semimetal}
Consider a general Hamiltonian written in form of Eq.~(\ref{HappD}). The inversion symmetric Hamiltonian in Eq.~(\ref{NLSMclass}) is particular choice for $d_x=v(\sqrt{k_x^2+k_y^2}-k_0$, $d_y=0$ and $d_z=v k_z$. 

 Note that in an inversion symmetric system we need an external electric field to break the inversion symmetry in order to find second order response in the system. Let us assume that the external electric field is in the $z$ direction. This external electric field changes the electron distribution function in conduction and valence band which we can write as following up to the first order in the external electric field 
\begin{equation}
    f_{n}=f_{n}^{(0)}+\frac{e\tau }{\hbar} E_z \partial_{k_z}\epsilon \frac{\partial f_{n}^{(0)}(\epsilon)}{\partial \epsilon(k)}\,.
\end{equation}
Here, $f_{n}^{(0)}(\epsilon)$ is the equilibrium distribution function, $n$ is the band index and $\epsilon$ is the conduction band's energy. 

There are four contributions to the SHG in Eq.~(\ref{SHG2pItwoband})-Eq.~(\ref{SHG1pIItwoband}) for two-band systems which we are going to investigate for Hamiltonian Eq.~(\ref{NLSMclass}). 

\begin{align}
\label{SHG2pItwobandappapp}
    &\sigma^{ijm}_{2p,I}(\omega)=\frac{e^3}{2\hbar^2 \omega^2}\int [dk] M_{2pI,cv}^{ijm}(k)f_{vc}R_{\gamma}(2\omega-\omega_{cv}),\\&
    \label{SHG2pIItwobandappapp}
    \sigma^{ijm}_{2p,II}(\omega)=\frac{ e^3}{2\hbar^2 \omega^2}\int [dk] M_{2pII,cv}^{ijm}(k)f_{vc}R_{\gamma}(2\omega-\omega_{cv}),\\&
        \label{SHG1pItwobandappapp}
    \sigma^{ijm}_{1p,I}(\omega)=\frac{e^3}{2\hbar^2 \omega^2}\int [dk] M_{1pI,cv}^{ijm}(k)f_{vc}R_{\gamma}(2\omega-\omega_{cv}),\\&
        \label{SHG1pIItwobandappapp}
    \sigma^{ijm}_{1p,II}(\omega)=\frac{ e^3}{2\hbar^2 \omega^2}\int [dk] M_{1pII,cv}^{ijm}(k)f_{vc}R_{\gamma}(2\omega-\omega_{cv}),
\end{align}
where, 
\begin{align}
\label{M2pIappapp}
    &M_{2pI,cv}^{ijm}=v^{i}_{vc}w_{cv}^{jm},\\&
    \label{M2pIIappapp}
    M_{2pII,cv}^{ijm}=\frac{-4v_{vc}^{i}\big[v_{cv}^{j},v_{cc}^{m}\big]_{+}}{\omega_{cv}},\\&
        \label{M1pIappapp}
        M_{1pI,cv}^{ijm}=w^{ij}_{vc}v_{cv}^{m}+w_{vc}^{im}v^{j}_{cv},\\&
        \label{M1pIIappapp}
     M_{1pII,cv}^{ijm}=\frac{2v_{vc}^{i}}{\omega_{cv}}\big[v_{cc}^{j},v_{cv}^{m}\big]_{+}-\frac{v_{cc}^{i}}{\omega_{cv}}\big[v_{cv}^{j},v_{vc}^{m}\big]_{+},
\end{align}
and $w^{ab}=(1/\hbar)\partial_{k_a}\partial_{k_b}H$, $[v^{b}_{np}v^{c}_{pm}]_{+}=v^b_{np}v^c_{pm}+v^c_{np}v^b_{pm}$ and $R_{\gamma}(x)=1/(x-i\gamma)$, where $\gamma$ is related to the decay rate.

Let us consider $|1\big>$ as the conduction and $|0\big>$ as the valence band. Using the following identity for Pauli matrices, $\tau_i$, we can find the general form of $M_I$ and $M_{II}$ for the $2\times2$ Hamiltonians ($H=\sum_{\alpha=1}^{3}d_{\alpha}\tau_{\alpha}$)
\begin{equation}
    \big<0|\tau_\alpha|1\big>\big<1|\tau_\beta|0\big>=(\delta_{ij}-\frac{d_\alpha d_{\beta}}{d^2})-i\epsilon_{\alpha\beta\gamma}\frac{d_\gamma}{d}\,,
\end{equation}
thus we have 
\begin{align}
\label{rel1}
    &v_{vc}^iv_{cv}^j=\frac{\partial d_{\alpha}}{\partial k_i}\frac{\partial d_{\beta}}{\partial k_j}\big<0|\tau_{\alpha}|1\big>\big<1|\tau_{\beta}|0\big>=\frac{\partial d_{\alpha}}{\partial k_i}\frac{\partial d_{\beta}}{\partial k_j}\big((\delta_{\alpha\beta}-\frac{d_\alpha d_{\beta}}{d^2})-i\frac{d_{\gamma}}{d}\epsilon_{\alpha\beta\gamma}\big)\,,\\&
    v_{vc}^iw_{cv}^{jm}=\frac{\partial d_{\alpha}}{\partial k_i}\frac{\partial^2 d_{\beta}}{\partial k_j\partial k_m}\big<0|\tau_{\alpha}|1\big>\big<1|\tau_{\beta}|0\big>=\frac{\partial d_{\alpha}}{\partial k_i}\frac{\partial^2 d_{\beta}}{\partial k_j\partial k_m}\big((\delta_{\alpha\beta}-\frac{d_\alpha d_{\beta}}{d^2})-i\frac{d_{\gamma}}{d}\epsilon_{\alpha\beta\gamma}\big)\,.
    \label{rel2}
\end{align}
Using Eq.~(\ref{rel1}) and Eq.~(\ref{rel2}) in Eq.~(\ref{M2pIappapp})-Eq.~(\ref{M1pIIappapp}), and seprating the imaginary and real part of each function we find 
\begin{align}
    &\operatorname{Re}[M_{2pI}^{ijm}]=\frac{\partial d_{\alpha}}{\partial k_i}\frac{\partial^2 d_{\beta}}{\partial k_j\partial k_m}(\delta_{\alpha\beta}-\frac{d_\alpha d_{\beta}}{d^2})\,,\\&
    \operatorname{Im}[M_{2pI}^{ijm}]=-\frac{\partial d_{\alpha}}{\partial k_i}\frac{\partial^2 d_{\beta}}{\partial k_j\partial k_m}\frac{d_{\gamma}}{d}\epsilon_{\alpha\beta\gamma}\,,
\end{align}
\begin{align}
    &\operatorname{Re}[M_{2pII}^{ijm}]=\frac{-2}{d}\frac{\partial d_{\alpha}}{\partial k_i}\big(\frac{\partial d_{\beta}}{\partial k_j}\frac{\partial d}{\partial k_m}+\frac{\partial d_{\beta}}{\partial k_m}\frac{\partial d}{\partial k_j}\big)(\delta_{\alpha\beta}-\frac{d_\alpha d_{\beta}}{d^2})\,,\\&
    \operatorname{Im}[M_{2pII}^{ijm}]=\frac{2}{d}\frac{\partial d_{\alpha}}{\partial k_i}\big(\frac{\partial d_{\beta}}{\partial k_j}\frac{\partial d}{\partial k_m}+\frac{\partial d_{\beta}}{\partial k_m}\frac{\partial d}{\partial k_j}\big)\frac{d_{\gamma}}{d}\epsilon_{\alpha\beta\gamma}\,,
\end{align}
\begin{align}
    &\operatorname{Re}[M_{1pI}^{ijm}]=\big(\frac{\partial d_{\beta}}{\partial k_m}\frac{\partial^2 d_{\alpha}}{\partial k_i\partial k_j}+(j \leftrightarrow m)\big)(\delta_{\alpha\beta}-\frac{d_\alpha d_{\beta}}{d^2})\,,\\&
    \operatorname{Im}[M_{1pI}^{ijm}]=-\big(\frac{\partial d_{\beta}}{\partial k_m}\frac{\partial^2 d_{\alpha}}{\partial k_i\partial k_j}+(j \leftrightarrow m)\big)\frac{d_{\gamma}}{d}\epsilon_{\alpha\beta\gamma}\,,
\end{align}
\begin{align}
    &\operatorname{Re}[M_{1pII}^{ijm}]=\frac{1}{d}\big(\frac{\partial d_{\alpha}}{\partial k_i}\frac{\partial d_{\beta}}{\partial k_m}\frac{\partial d}{\partial k_j}+(j \leftrightarrow m)\big)(\delta_{\alpha\beta}-\frac{d_\alpha d_{\beta}}{d^2})-\frac{1}{d}\frac{\partial d}{\partial k_i}\frac{\partial d_{\beta}}{\partial k_m}\frac{\partial d_{\alpha}}{\partial k_j}(\delta_{\alpha\beta}-\frac{d_\alpha d_{\beta}}{d^2})\,,\\&
    \operatorname{Im}[M_{1pII}^{ijm}]=\frac{-1}{d^2}\big(\frac{\partial d_{\alpha}}{\partial k_i}\frac{\partial d_{\beta}}{\partial k_m}\frac{\partial d}{\partial k_j}+(j \leftrightarrow m)\big)d_{\gamma}\epsilon_{\alpha\beta\gamma}\,.
\end{align}
Note that there is a sum on repeated indexes ($\alpha$,$\beta$, and $\gamma$). 
It is easy to see because $d_y=0$, all the imaginary parts vanish in the inversion symmetric NLSM Hamiltonian Eq.~(\ref{NLSMclass}). At low frequency limit the contribution of $M_{2pI}^{ijm}$ and $M_{1pI}^{ijm}$ to the SHG vanish and only $M_{2pII}$ and $M_{1pII}$ have finite contribution. 
Let us consider the $zzz$ direction. In this direction, $M_{1pI}^{zzz}$ and $M_{2pI}$ vanish explicitly because the Hamiltonian is linear in $k_z$. For $M_{2pII}$ and $M_{1pII}$ we have the following expressions 
\begin{equation}
    \operatorname{Re}[M_{2pII}^{zzz}]=\frac{-4(\sqrt{k_x^2+k_y^2}-k_0)^2k_zv^2}{\big((\sqrt{k_x^2+k_y^2}-k_0)^2+k_z^2\big)^2}\,.
    \label{M2pIIzzzapp}
\end{equation}
Also it can be seen $Re[M_{1pII}^{zzz}]=-\frac{1}{4}Re[M_{2pII}^{zzz}]$. Now let us use the following variable change
\begin{align}
\label{varchangeSHG}
    &k_x= (\frac{d\sin(\theta)}{v}+k_0)\sin(\phi)\,,\\\nonumber&
    k_y= (\frac{d\sin(\theta)}{v}+k_0)\cos(\phi)\,,\\\nonumber&
    k_z= \frac{d \cos(\theta)}{v}\,,
\end{align}
where the determinant of the above transformation is given by 
\begin{equation}
    \det|J|=\frac{d(vk_0+d\sin(\theta))}{v^2}\,.
\end{equation}
where $d=v\sqrt{(\sqrt{k_x^2+k_y^2}-k_0)^2+k_z^2}$. Using the above variable change we can simplify Eq.~(\ref{M2pIIzzzapp}) as following
\begin{equation}
    \operatorname{Re}[M_{2pII}^{zzz}]=\frac{-4v^3\cos(\theta)\sin(\theta)^2}{d}\,.
    \label{i}
\end{equation}
inserting Eq.~(\ref{i}) in Eq.~(\ref{SHG2pIItwobandappapp}) we can find 
\begin{equation}
    \sigma_{2p,II}^{zzz}=\frac{-e^4E_zv\tau}{2h^3\omega^2}\int_{0}^{\infty} dd\int_{0}^{2\pi}d\theta \int_{0}^{2\pi}d\phi (k_0v+d\sin(\theta))\sin(2\theta)^2R_{\gamma}(2\omega-2d)\delta(d-\mu) 
\end{equation}
The only term that have a finite contribution, is the term proportional to $k_0$ and the rest vanish due to the angular integral. Thus we find 
\begin{equation}
    \sigma_{2p,II}^{zzz}=-\frac{e^4E_zk_0\pi^2v^2\tau}{h^3\omega^2}R_{\gamma}(2\omega-2\mu)\,.
\end{equation}
Also by using $Re[M_{1pII}^{zzz}]=-\frac{1}{4}Re[M_{2pII}^{zzz}]$, for the $\sigma_{1pII}^{zzz}$ we can find the following expression
\begin{equation}
    \sigma_{1p,II}^{zzz}=\frac{e^4E_zk_0\pi^2v^2\tau}{4h^3\omega^2}R_{\gamma}(\omega-2\mu) 
\end{equation}

In the $zxx$ direction, $M_{1pI}^{zxx}$ vanishes explicitly but the $M_{2pI}$ is not explicitly zero. However, the angular integral makes this term vanish. Thus, the only contributors come from $M_{1pII}^{zxx}$ and $M_{2pII}^{zxx}$. By changing the variables Eq.~(\ref{varchangeSHG}) we can simply write two contributions as following
\begin{align}
        &\sigma_{2p,II}^{zxx}=\frac{e^4E_zv\tau}{2h^3\omega^2}\int_{0}^{\infty} dd\int_{0}^{2\pi}d\theta \int_{0}^{2\pi}d\phi (k_0v+d\sin(\theta))\sin(2\theta)^2\sin(\phi)^2R_{\gamma}(2\omega-2d)\delta(d-\mu)\,,\\&
        \sigma_{1p,II}^{zxx}=\frac{e^4E_zv\tau}{2h^3\omega^2}\int_{0}^{\infty} dd\int_{0}^{2\pi}d\theta \int_{0}^{2\pi}d\phi (k_0v+d\sin(\theta))\cos(\theta)^2(-3+\cos(2\theta))\sin(\phi)^2R_{\gamma}(\omega-2d)\delta(d-\mu)\,.
\end{align}
By evaluating the integrals we find 
\begin{align}
    &\sigma_{2p,II}^{zxx}=\frac{e^4E_z\tau\pi^2k_0v^2}{2h^3\omega^2}R_{\gamma}(2\omega-2\mu)\,,\\&
    \sigma_{1p,II}^{zxx}=\frac{-5e^4E_z\tau\pi^2k_0v^2}{8h^3\omega^2}R_{\gamma}(\omega-2\mu)\,.
\end{align}

The same approach can be used to find the SHG in $xxz$ and $xzx$ directions, in which we find the following expressions 
\begin{align}
    &\sigma_{2p,II}^{xxz}=\sigma_{2p,II}^{xzx}=-\frac{e^4E_z\tau\pi^2k_0v^2}{2h^3\omega^2}R_{\gamma}(2\omega-2\mu)\,,\\&
    \sigma_{1p,II}^{zxx}=\sigma_{1p,II}^{xzx}=\frac{3e^4E_z\tau\pi^2k_0v^2}{8h^3\omega^2}R_{\gamma}(\omega-2\mu)\,.
\end{align}

\begin{table}[]
    \begin{tabular}{ |p{1.3cm}|p{1.3cm}|p{1.3cm}|  }

    \hline
    \multicolumn{3}{|c|}{Coefficients of 1pII and 2pII resonances} \\
    \hline
    abc & $C_{1pII}^{abc}$ & $C_{1pII}^{abc}$ \\
    \hline

    zzz & $-\pi^2$ & $\frac{\pi^2}{16}$ \\ 
    zxx & $\frac{\pi^2}{2}$ & $\frac{-5 \pi^2}{32}$ \\
    xzx & $\frac{-\pi^2}{2}$ & $\frac{3\pi^2}{32}$ \\
    xxz & $\frac{-\pi^2}{2}$ & $\frac{3\pi^2}{32}$ \\
    \hline
    \end{tabular}
    \caption{The coefficients of Eq.~(\ref{Eq_results_SHG}) for different directions ($a,b,c\in \{x,z\}$). Note that the other four directions (zzx, zxz, xzz, xxx) that is not included in the table vanish.}
    \label{coeftab}
\end{table}

\section{}
\subsection{Definitions}

{\bf Hamiltonian}:
\begin{equation}
    H(k)=\sum_{\alpha=1}^{3}d_{\alpha}(k)\sigma_{\alpha}\,,
    \label{HappD}
\end{equation}
where $\sigma_{1}=\sigma_{x}$, $\sigma_{2}=\sigma_{y}$ and $\sigma_{3}=\sigma_{z}$ are Pauli Matrices. 

{\bf shift conductivity}:
\begin{equation}
    \sigma_{shift}^{ijm}(0;\omega,-\omega)=\frac{2\pi e^3}{\hbar^2}\int [dk] I^{ijm}_{10} f_{01}\delta(\omega-\omega_{10})\,,
\end{equation}
where $1$ and $0$ denote conduction and valence band, respectively and for tensor $I_{10}^{ijm}$ we have (we ignore the index 10 for the $I^{ijm}_{10}$ tensor for simplicity)
\begin{equation}
   I^{ijm}=\frac{-1}{2}\operatorname{Im}\big[r_{10}^jr^m_{01;i}+r_{10}^mr^j_{01;i}\big]\,.
   \label{Iapprnm}
\end{equation}
Here $r^m_{01;i}=\partial_{k_i}r^m_{01}-i(\xi_{00}^i-\xi_{11}^i)r^m_{01}$ is the generalized derivative. To see the effect of berry curvature we can expand the $I$ tensor
\begin{equation}
    I^{ijm}=\frac{1}{2}|r^{j}_{10}||r^{m}_{01}|\big(\partial_{k_i}(\phi^j+\phi^m)-2(\xi^i_{00}-\xi_{11}^i)\big)\cos(\phi^j-\phi^m)+\frac{1}{2}|r_{10}^j|^2\partial_{k_i}(\frac{|r_{01}^m|}{|r_{10}^j|})\sin(\phi^j-\phi^m)\,,
\end{equation}
where, $\phi^j=\arg(r_{10}^j)$. We can see when $j$ and $m$ are in the same direction, then the $\sin$ term vanish and the first term become the shift vector definition $R^i_{10,j}=\partial_{k_i}\phi^j-(\xi_{00}^i-\xi_{11}^i)$.

We can define the $I^{ijm}$ in terms of the Hamiltonian terms as following 
\begin{equation}
      I^{ijm}_{10}=\frac{-1}{8d^3}\sum_{\alpha\beta\gamma}\big[d_{\gamma}d_{\alpha,j}d_{\beta,im}-d_{\alpha,j}d_{\beta,i}d_{\gamma}\frac{d,m}{d}+(j\leftrightarrow m)\big]\epsilon_{\alpha\beta\gamma}
      \label{Iappdijk}\,,
\end{equation}
or equivalently we can write the $I^{ijm}$ tensor as following 
\begin{equation}
    I^{ijm}=\frac{1}{\omega_{10}^2}\operatorname{Im}\bigg(\frac{-v_{10}^i\big[v_{01}^jv_{11}^m\big]_+}{\omega_{10}}+\frac{1}{2}(v_{01}^mw_{10}^{ij}+v_{01}^jw^{im}_{10})\bigg)\,.
    \label{Iappvnm}
\end{equation}

{\bf Proof 1}:\\
In the proof 1 we show that Eq.~(\ref{Iapprnm}) is equal to Eq.~(\ref{Iappvnm})
To find the $r_{nm}^i$ we can start with taking the derivative of the Hamiltonian  
\begin{equation}
    \partial_{k_i}\big<n|H|m\big>= \epsilon_{m}\big<\partial_{k_i}n|m\big>+\big<n|\partial_{k_i}H|m\big>+\epsilon_{n}\big<n|\partial_{k_i}m\big>=0\,\,\,\,\,\,\,\,(\text{for} n \neq m)\,.
\end{equation}
By using the fact that $\big<\partial_{k_i}n|m\big>=-\big<n|\partial_{k_i}m\big>$ and using the definition that $r_{nm}=i\big<n|\partial_{k_i}m\big>$ and $v_{nm}^i=\frac{1}{\hbar}\big<n|\partial_{k_i}H|m\big>$ we can find 
\begin{equation}
i r_{nm}=\frac{v_{nm}}{\omega-\omega_m}\,.
\label{rnmapp}
\end{equation}
Now we can take the derivative of the Eq.~(\ref{rnmapp}) in which we have 
\begin{equation}
    ir_{nm;i}^{j}=\frac{v_{nm;i}^j}{\omega_{nm}}-\frac{v^j_{nm}}{\omega_{nm}^2}\partial_{k_i}\omega_{nm}\,.
    \label{covrapp}
\end{equation}
We can define the derivative of the element of the velocity operator as following
\begin{equation}
    \partial_{k_i}v_{nm}=\partial_{k_i}(\big<n|\hat{v}|m\big>)=\big<\partial_{k_i}n|\hat{v}|m\big>+\big<n|\partial_{k_i}\hat{v}|m\big>+\big<n|\hat{v}|\partial_{k_i}m\big>\,.
\end{equation}
By inserting identity operator we can find
\begin{equation}
    \partial_{k_i}v^j_{nm}=w_{nm}^{ij}+\big<\partial_{k_i}n|n\big>v_{nm}^j+\sum_{p\neq n}\big<\partial_{k_i}n|p\big>v_{pm}^j+\sum_{p\neq m
    }v_{np}^j\big<p|\partial_{k_i}m\big>+v_{nm}^j\big<m|\partial_{k_i}m\big>\,.
\end{equation}
Thus we have 
\begin{equation}
    \partial_{k_i}v^j_{nm}=w_{nm}^{ij}+i(\xi_{nn}^i-\xi_{mm}^j)v_{nm}^j-i\sum_{p\neq m
    }v_{np}^j\xi_{pm}^i+i\sum_{p\neq n}\xi_{np}^{i}v_{pm}^j\,.
\end{equation}
Using the definition of generalized derivative we have 
\begin{equation}
    v_{nm;i}^j= \partial_{k_i}v^j_{nm}-i(\xi_{nn}^i-\xi_{mm}^i)v_{nm}^j=w_{nm}^{ij}-i\sum_{p\neq m
    }v_{np}^j\xi_{pm}^i+i\sum_{p\neq n}\xi_{np}^{i}v_{pm}^j\,,
\end{equation}
also, using the fact that $ir^i_{nm}=\frac{v^i_{nm}}{\omega_{nm}}$ for $n\neq m$ for a two-band systems we have 
\begin{equation}
    v_{01;i}^{j}=w_{01}^{ij}+\frac{v_{01}^i(v_{11}^j-v_{00}^j)}{\omega_{01}}\,.
    \label{covV}
\end{equation}
Inserting Eq.~(\ref{covV}) in Eq.~(\ref{covrapp}) for a two-band system, we have
\begin{equation}
    i\omega_{01}r_{01;i}^{j}=w_{01}^{ij}+\frac{v_{01}^j(v^i_{11}-v_{00}^i)}{\omega_{01}}+\frac{v_{01}^i(v^j_{11}-v_{00}^j)}{\omega_{01}}\,.
    \label{covr_twoband}
\end{equation}

Finally by inserting Eq.~(\ref{covr_twoband}) in Eq.~(\ref{Iapprnm}) we can find Eq.~(\ref{Iappvnm}).

{\bf Proof 2:}
In proof 2 we show that \ref{Iappvnm} is equal to \ref{Iappdijk}
let us consider only the following part 
\begin{equation}
    \frac{1}{\omega_{10}^2}\operatorname{Im}\big[\frac{-v_{10}^iv_{01}^jv_{11}^m}{\omega_{10}}+\frac{1}{2}v_{01}^jw^{im}_{10}\big]\,.
\end{equation}
The other terms will be generated by $(j \leftrightarrow m)$.
We can simplify the above expression using the definition of $v_{nm}^i=\frac{1}{\hbar}\big<n|\partial_{k_i}H|m\big>$,  $v_{nm}^i=\frac{1}{\hbar}\big<n|\partial_{k_i}\partial_{k_j}H|m\big>$ and $\omega_{10}=2d/\hbar$ in which we find 
\begin{equation}
    \frac{1}{4d^2}\operatorname{Im}\big[\frac{-1}{2d}\partial_{k_i}d_{\alpha}\partial_{k_j}d_{\beta}\big<1|\sigma_{\alpha}|0\big>\big<0|\sigma_{\beta}|1\big>\partial_{k_m}d_\rho\big<1|\sigma_{\rho}|1\big>+\frac{1}{2}\partial_{k_m}d_{\alpha}\partial_{k_i}\partial_{k_j}d_{\beta}\big<0|\sigma_{\alpha}|1\big>\big<1|\sigma_{\beta}|0\big>\big]\,.
    \label{EQDcom}
\end{equation}
Using the following identities we can simplify the above equation 
\begin{align}
    &\big<1|\sigma_{\alpha}|0\big>\big<0|\sigma_{\beta}|1\big>=(\delta_{\alpha\beta}-\frac{d_{\alpha}d_{\beta}}{d^2})+i\epsilon_{\alpha\beta\gamma}\frac{d_{\gamma}}{d}\\
    &\big<1|\sigma_{\alpha}|1\big>=\frac{d_{\alpha}}{d}\,.
\end{align}
Here, $\epsilon_{\alpha\beta\gamma}$ is the Levi-Civita symbol.
Now by considering the imaginary part of the Eq.~(\ref{EQDcom}) we find 
\begin{equation}
    \frac{1}{4d^2}\big[-\partial_{k_i}d_\alpha\partial_{k_j}d_\beta \frac{d_ld_{\gamma}}{2d^2}\partial_{k_m}d_l-\partial_{k_m}\partial_{k_i}d_{\beta}\partial d_{\alpha}\frac{d_{\gamma}}{2d}\big]\epsilon_{\alpha\beta\gamma}\,.
\end{equation}
By changing the $\alpha$ and $\beta$ index in the first term and using the identity $d_l\partial_{k_m}d_l=d\partial_{k_m}d$ we can find 
\begin{equation}
      \frac{-1}{8d^3}\sum_{\alpha\beta\gamma}\big[d_{\gamma}d_{\alpha,j}d_{\beta,im}-d_{\alpha,j}d_{\beta,i}d_{\gamma}\frac{d,m}{d}\big]\epsilon_{\alpha\beta\gamma}\,,
      \label{Iappdijk}
\end{equation}
which is the first term in the Eq.~(\ref{Iappdijk}).

\subsection{SHG and shift current relation}
Let us look at the general case of $\sigma_{SHG}^{ijj}$. Because of the symmetry of the Hamiltonian between $x$ and $y$ in our model, we consider $i$ and $j$ to be $z$ or $x$ only. For the SHG response in the $ijj$ directoin we have 
\begin{align}
\label{M2pIijjapp}
    &M_{2pI}^{ijj}=v_{01}^{i}w_{10}^{jj}\,,\\
    \label{M2pIIijjapp}
    &M_{2pII}^{ijj}=\frac{-8v_{01}^iv_{10}^jv_{11}^j}{\omega_{10}}\,,\\
    \label{M1pIijjapp}
    &M_{1pI}^{ijj}=2w_{01}^{ij}v_{10}^j\,,\\
    \label{M1pIIijjapp}
    &M_{1pII}^{ijj}=4\frac{v_{01}^iv_{10}^jv_{11}^j}{\omega_{10}}-\frac{2v_{11}^iv_{10}^jv_{01}^j}{\omega_{10}}\,.
\end{align}
Here, $0$ and $1$ denote valence and conduction band, respectively.
Now we consider two cases. First, the case that $i\neq j$ and second $i=j$. For the first case, in our system, $w^{ij}_{01}$ vanishes because there is no crossing term such as $k_xk_z$ in the Hamiltonian. Also the imaginary part of the second term in Eq.~(\ref{M1pIijjapp}) vanishes because it is fully real. 

Now let us look at the shift conductivity in the $ijj$ and $jij$ directions.
\begin{equation}
    \omega_{10}^2I^{ijj}=\operatorname{Im}\big[\frac{-2v_{10}^iv_{01}^jv_{11}^j}{\omega_{10}}+v_{01}^jw_{10}^{ij}\big]\,,
    \label{Iijjapp}
\end{equation}
\begin{equation}
    \omega_{10}^2I^{jij}=\operatorname{Im}\big[\frac{-v_{10}^jv_{01}^iv_{11}^j}{\omega_{10}}-\frac{v_{10}^jv_{01}^jv_{11}^i}{\omega_{10}}+\frac{1}{2}v_{01}^jw_{10}^{ij}+\frac{1}{2}v_{01}^iw_{10}^{jj}\big]\,.
\label{Ijijapp}
\end{equation}
Here, we can see the second term in Eq.~(\ref{Ijijapp}) is completely real, and the third term vanishes because, in our system, we do not have crossing terms ($i \neq j$). Thus we can see we can simplifty the above equations as following 
\begin{equation}
    \omega_{10}^2I^{ijj}=\frac{-1}{4}\operatorname{Im}\big[M_{2pII}^{ijj}\big]=\frac{1}{2}\operatorname{Im}\big[M_{1pII}^{ijj}\big]=\operatorname{Im}\big[\frac{-v_{10}^jv_{01}^iv_{11}^j}{\omega_{10}}\big]\,,
    \label{ineqjSHGshift1}
\end{equation}
\begin{equation}
    \operatorname{Im}\big[M_{2pI}^{ijj}\big]=\operatorname{Im}\big[v_{01}^iw_{10}^{jj}\big]=2\omega_{10}^2I^{jij}+\omega^2_{10}I^{ijj}\,,
    \label{ineqjSHGshift2}
\end{equation}
Now we are able to find the SHG as following 
\begin{equation}
    \sigma_{SHG}^{ijj}(2\omega;\omega,\omega)=\frac{i\pi e^3}{2\hbar^2\omega^2}\int[dk]\bigg( (M_{2pII}^{ijj}+M_{2pI}^{ijj})\delta(2\omega-\omega_{10})+M_{1pII}^{ijj}\delta(\omega-\omega_{10})\bigg)f_{01}\,.
\end{equation}
Using Eq.~(\ref{ineqjSHGshift1}) and Eq.~(\ref{ineqjSHGshift2}) we can find that 
\begin{equation}
    \operatorname{Re}\big[\sigma_{SHG}^{ijj}(2\omega;\omega,\omega)\big]=\frac{\pi e^3}{2\hbar^2\omega^2}\int[dk]\bigg(\omega_{10}^2(-3I^{ijj}+2I^{jij})\delta(2\omega-\omega_{10})+2\omega_{10}^2I^{ijj}\bigg)f_{01}\,,
\end{equation}
which we can simplify the $\omega^2$ term in the denominator using the delta function (note that it gives a factor of 4 for the 2photon resonance processes)
\begin{equation}
    \operatorname{Re}\big[\sigma_{SHG}^{ijj}(2\omega;\omega,\omega)\big]=\frac{\pi e^3}{\hbar^2}\int[dk]\bigg(2(-3I^{ijj}+2I^{jij})\delta(2\omega-\omega_{10})+I^{ijj}\bigg)f_{01}\,,
\end{equation}
thus by the definition of the shift conductivity we have  
\begin{equation}
    \operatorname{Re}\big[\sigma_{SHG}^{ijj}(2\omega;\omega,\omega)\big]=-3\sigma_{shift}^{ijj}(0;2\omega,-2\omega)+2\sigma_{shift}^{jij}(0;2\omega,-2\omega)+\frac{1}{2}\sigma_{shift}^{ijj}(0;\omega,-\omega)\,.
\end{equation}

For the case that $i=j$ Eq.~(\ref{M2pIijjapp}) and Eq.~(\ref{M1pIIijjapp}) vanish because both are real. For the remaining terms we have 
\begin{equation}
   \operatorname{Im}\big[ M_{2pI}^{iii}\big]=-\frac{1}{2}\operatorname{Im}\big[M_{1pI}^{iii}\big]=\omega_{10}^2I^{iii}\,.
   \label{Msrelation}
\end{equation}
Thus we have 
\begin{equation}
    \operatorname{Re}\big[\sigma_{SHG}^{iii}(2\omega;\omega,\omega)\big]=\frac{\pi e^3}{2\hbar^2\omega^2}\int[dk]\bigg(\omega_{10}^2I^{iii}\delta(2\omega-\omega_{10})-2\omega_{10}^2I^{iii}\delta(\omega-\omega_{10})\bigg)f_{01}\,,
\end{equation}
we can simplify the $\omega^2$ term in the denominator using the delta function in which we find 
\begin{equation}
    \operatorname{Re}\big[\sigma_{SHG}^{iii}(2\omega;\omega,\omega)\big]=\sigma_{shift}^{iii}(0;2\omega,-2\omega)-\frac{1}{2}\sigma_{shift}^{iii}(0;\omega,-\omega)\,.
\end{equation}



\end{document}